\documentstyle[prl,aps]{revtex}
\input epsf

\begin{document}
 
\newcommand{\bi}{\begin{itemize}}  
\newcommand{\ei}{\end{itemize}} 
\newcommand{\ii}{\item} 
\newcommand{\f}{\frac}
\newcommand{\la}{\label}
\def\be{\begin{equation}}
\def\ee{\end{equation}}

\title{\bf On positive functions with positive Fourier transforms}

\author{B.G. Giraud\thanks{giraud@dsm-mail.saclay.cea.fr} and R. 
Peschanski\thanks{pesch@dsm-mail.saclay.cea.fr}}
\address{Service de 
Physique Th\'eorique, DSM, CE Saclay, F-91191 Gif/Yvette, France}

\maketitle

\begin{abstract}
Using the basis of Hermite-Fourier functions (i.e. the quantum oscillator 
eigenstates) and the Sturm theorem, we derive the practical constraints for 
a function and its Fourier transform to be both positive. We propose a 
constructive method based on the algebra of Hermite polynomials. 
Applications are extended to  the 2-dimensional case  (i.e. 
Fourier-Bessel transforms and the algebra of Laguerre polynomials) and to 
adding constraints on derivatives, such as monotonicity or convexity.
\end{abstract}

\section{Introduction}

Positivity conditions for a Fourier transform appear in various domains of 
Physics. Frequent questions are, 
\bi
\ii 
What are the constraints for a real function $\psi(r)$  ensuring that its 
Fourier transform,
\be
\varphi(s)=\frac{1}{\sqrt{2\pi}}\, \int_{-\infty}^{+\infty} dr\ e^{i s r}\, 
\psi(r)\, ,
\ee
\noindent be real and positive? 
\ii 
Conversely, what are the properties of $\varphi$ if $\psi$ is positive? 
\ii 
Finally, what are the constraints on Fourier partners such that  
{\it both} $\psi$ and $\varphi$ be positive?
\ei
Physicists often need to work with concrete constructions. The practical 
construction of a basis of  functions satisfying the abovementionned 
positivity properties remains, up to our knowledge, an 
open problem. Such questions are quite relevant in Physics. As practical 
examples, let us quote two well-known cases. A Fourier transform 
relates \cite{Glau} two quantities, namely the cross section and the profile 
of a nucleus, which ought to be both positive. In Particle Physics, a 2-d 
Fourier-Bessel transform relates the color dipole distribution in transverse 
position space (derived from Quantum Chromodynamics) and the transverse 
momentum distribution of  gluons probed during  a deep-inelastic collision 
\cite{Kovchegov}. In short, such questions occur also in probability calculus,
for the relation between probability distributions and characteristic 
functions \cite{levy}, in crystallography and more generally in condensed 
matter physics, e.g. for the interpretation of patterns, etc....

The problem is simplified if related to another one, that
concerning the functions which are invariant \cite{Titch} up to a phase 
factor\footnote{They are called ``self-Fourier'' in \cite{Caola:1991ad} 
if the phase factor is 1 and ``generalized self-Fourier'' 
\cite{Cincotti:1991ad} (or ``dual'') for other phases.} 
by Fourier transforms. Indeed,  the most familiar examples of positive 
self-dual functions or distributions, thus trivially  verifying the double 
positivity condition,  which are at most scaled under Fourier 
transformation (FT), are the Gaussian and the Dirac comb. Many special 
cases can be found, where positivity is conserved, such as, for 
instance, the continuous family of functions $\exp(-r^\nu),$ where 
$0<\nu \le 2.$ Various {\it sufficient} conditions for positivity
can be found in the literature, such as the convexity of $\psi$ 
\cite{lafforgue} but, up to our knowledge, no general constructive method 
has been presented. 

The present note attempts to give general positivity criteria, in a 
constructive way, by taking advantage of a representation under which the 
FT is essentially ``transparent''. Our method combines the advantages 
of self-duality properties with those allowed by  an algebra of polynomials, 
where positivity means absence of real roots, hence explicit manipulations 
of polynomial coefficients. For this sake, in the 1-d  case, we select a 
basis made of convenient eigenstates of the FT, the Hermite-Fourier 
functions, i.e. the harmonic oscillator eigenstates. The method extends 
to the 2-d case, or Fourier-Bessel transform, by replacing 
Hermite by Laguerre polynomials.

There are general mathematical theorems about the characterization of Fourier 
transforms of positive 
functions \cite{Scharnhorst:2002im}. Let us quote in the first place  the 
Bochner theorem and its generalizations \cite{gelfand} which state that the 
Fourier transform of a positive function is {\it positive-definite}. But
positive definiteness in the sense of such theorems does not imply plain
positivity\footnote{Positive definiteness means that for any real numbers 
$x_1...x_k$ and complex numbers $\xi_1...\xi_k,$ one has 
$\sum_{k,j}\varphi(x_k\!-\!x_j)\, {\bar \xi_j} \xi_k\ge 0.$}. Hence our 
problem actually could be rephrased \cite{Teugels:1971ge} as ``build 
positive-definite functions 
that are positive''.

Our formalism is the subject of Section II. Numerical, 
illustrative examples will be given in Section III. Then Section IV 
describes the extension of our algorithms to the 2-d problem. 
Hermite polynomials will be replaced by Laguerre ones, 
but the algebra remains essentially the same. A brief discussion, 
conclusion and outlook make Section V.

\section{Basic formalism}

Consider the harmonic oscillator Hamiltonian, $\frac{1}{2}(p^2+r^2),$ 
and its eigenwavefunctions, 
\be 
u_n(r)=\pi^{-\frac{1}{4}}\, e^{-\frac{1}{2}r^2} H_n(r).
\ee
Here, we set $H_n$ to be a {\it square normalized} Hermite polynomial, with a 
positive coefficient for its highest power term. For the sake of clarity, we 
list the first polynomials as, $H_0=1,\, H_1=\sqrt{2}\ r,\, 
H_2=(2r^2-1)/\sqrt{2},\, H_3=(2r^3-3r)/\sqrt{3}\, $ and their recursion 
relation
\be
a_{n+1}H_{n+1}=2\, r\,  a_{n}H_{n} -2\, n\, a_{n-1}H_{n-1}\ ,
\la{Hermite}
\ee
where $a_{n}= \sqrt{2^n n!}\ .$ It is known that the FT of such states brings 
only a phase,
\begin{equation}
\frac{1}{\sqrt{2\pi}}\, \int_{-\infty}^{\infty} dr\ e^{i s r}\, u_n(r) = 
i^n\ u_n(s)\, ,
\label{transpa1}
\end{equation}
and thus such states give generalized self-dual functions with phase $i^n.$ 
If one expands $\psi$ in the oscillator basis, 
$\psi(r)=\sum_{n=0}^N \psi_n\, u_n(r),$ with a truncation at some degree $N,$ 
then all odd components $\psi_{2p+1}$ must vanish if $\varphi$ must be real, 
and the even rest splits, under FT, into an invariant part and a part with 
its sign reversed, namely
\begin{equation}
\varphi(s)=\pi^{-\frac{1}{4}}\, e^{-\frac{1}{2}s^2}\ [P_+(s)-P_- (s)],\ \ 
P_+(s)=\sum_{p=0}^{[N/4]} \psi_{4p}\, H_{4p}(s),\ \ 
P_- (s)=\sum_{p=0}^{[(N-2)/4]} \psi_{4p+2}\, H_{4p+2}(s)\, ,
\label{transfo1}
\end{equation}
where the usual symbols $[N/4]$ and $[(N-2)/4]$ mean, respectively, the 
entire parts of $N/4$ and $(N-2)/4.$

Notice that, when all components $\psi_n$ vanish except $\psi_0,$ then 
both $\psi$ and $\varphi$ are positive, because $H_0=1.$ Hence, 
one may, starting from this special point in the functional 
space of functions, investigate those domains of parameters $\psi_n$ 
where the polynomials ${\cal P}=P_++P_-$ and ${\cal Q}=P_+-P_-$ 
have no real root. Notice that only even powers of $r$ and $s$ 
are involved. It will therefore be convenient to use auxiliary variables 
such as $\rho=r^2$ and $\sigma=s^2,$ and the domain of interest for the 
parameters $\psi_n$ will correspond to the absence of real {\it positive} 
roots for both $\rho$ and $\sigma.$ 

The second ingredient of our approach is the well-known Sturm theorem 
\cite{sturm}  which gives the efficient way to characterize and localize the 
real roots of any given polynomial. The {\it Sturm criterion} can be expressed 
as follows:

\medskip \noindent
`` Given a polynomial ${\cal P}(x),$ its {\it Sturm sequence} 
${\cal S}(x) \equiv \left\{{\cal S}_1,{\cal S}_2,...\ {\cal S}_m,...
{\cal S}_{j\le N}\right\}$
is the set of polynomials
\be
{\cal S}_1={\cal P},\ {\cal S}_2=d{\cal P}/dx,\ ...\ {\cal S}_m
=-{\cal S}_{m-2}+\left[\f{{\cal S}_{m-2}}{{\cal S}_{m-1}}\right]\, {\cal 
S}_{m-1}\ ...\ ,
\la{sturmsequence}
\ee
where $\bf [\frac{\ }{\ }]$ designates the polynomial 
quotient\footnote{The Sturm sequence is thus made of polynomial remainders, 
with $(-)$ signs. It obviously stops at some $j \le N.$}. To know the number of 
roots between $x=a$ and $x=b,$ 
count the number ${\cal N}(a)$ of sign changes in ${\cal S}(a)$ and, 
similarly, count ${\cal N}(b).$ Then the number of roots is 
$\left|{\cal N}(b)-{\cal N}(a)\right|.$ ''

\medskip \noindent
The domain  borders where the root number, 
$\left|{\cal N}(+\infty)-{\cal N}(0)\right|,$ changes have to do with 
cancellations of the {\it resultant} $\cal R$ between ${\cal P}$ and 
$d{\cal P}/dx.$ The cancellation of ${\cal R}$ corresponds to collisions 
between conjugate complex roots becoming real roots and conversely. 
Because of the demanded positivity of $\rho$ and $\sigma,$
the borders have also to do with sign changes of ${\cal P}(0)$ and 
${\cal Q}(0),$ meaning real roots $\rho$ and $\sigma$ going through $0.$ 
All such technicalities are taken care of by the Sturm criterion, which 
furthermore allows the labeling of each domain by its precise number 
of real roots.

This will be implemented here in an explicit way, analytically 
as much as possible, then numerically and graphically, for a few cases of a 
general illustrative value. For this, we will plot the shapes of domains 
labeled by the values of the Sturm criterion. Hence the combination of the 
self-dual properties of the quantum oscillator basis and of the Sturm 
theorem allows a constructive method for a systematic investigation of 
positivity conditions for a 1-d Fourier transform. This basis has the 
potential to represent any function in the Hilbert space ${\cal L}^2$, but 
the present study concerns a finite set of components. An extension to 
infinite series of components may deserve other tools.

\section{Positivity domains}

In this Section we will apply our method to the 
case of a basis formed with 3 or 4 Hermite-Fourier functions. We consider: 
{\bf A} the basis ${\psi_0, \psi_4, \psi_8},$  then 
{\bf B} the basis ${\psi_0, \psi_4, \psi_8, \psi_{12}},$ which both lie  
in the subspace with eigenvalue $1$, and 
{\bf C} the basis ${\psi_0, \psi_2, \psi_4}$ where $\psi_2$ is in the 
subspace with eigenvalue $-1,$ furthermore, 
{\bf D} the influence of an additional constraint motivated by Physics,
that of monotony for $\varphi,$ and finally
{\bf E} a comparison with the convexity constraint for $\psi.$ Many other 
illustrations are possible, but these, {\bf A}-{\bf E}, demonstrate the 
flexibility of our approach.

\subsection{Mixture of three polynomials in the subspace with eigenvalue 1} 

Here we assume that $\psi$ has only components $\psi_0,$ $\psi_4$ and 
$\psi_8,$
hence ${\cal P}$ reduces to $P_+$ in (\ref{transfo1}) and we can study
\begin{equation}
{\cal P} = \psi_0 + \frac{\psi_4\,}{2\, \sqrt{6}}\,  (4 \rho^2-12 \rho+3)  + 
\frac{\psi_8\,}{24\, \sqrt{70}}
\, (16 \rho^4-224 \rho^3+840 \rho^2-840 \rho+105\,) ,
\end{equation}
where $\rho=r^2.$ One maintains $\psi_8 >0$ and one reads 
$P_+(0)=\psi_0+3 \psi_4/(2\, \sqrt{6})+105 \psi_8/(24\, \sqrt{70}).$ The 
resultant ${\cal R}$ between $P$ and $dP/d\rho$ is,
\begin{eqnarray}
{\cal R} \propto 
2100\, \psi_0\, \psi_4^4 - 1050\, \sqrt{6}\, \psi_4^5 - 
240\, \sqrt{70}\, \psi_0^2\, \psi_4^2\, \psi_8 + 
400\, \sqrt{105}\, \psi_0\, \psi_4^3\, \psi_8 - 
165\, \sqrt{70}\, \psi_4^4\, \psi_8 + 
\nonumber \\
480\, \psi_0^3\, \psi_8^2 + 
4560\, \sqrt{6}\, \psi_0^2\, \psi_4\, \psi_8^2 -
13320\, \psi_0\, \psi_4^2\, \psi_8^2 - 
1600\, \sqrt{6}\, \psi_4^3\, \psi_8^2 - 
792\, \sqrt{70}\, \psi_0^2\, \psi_8^3 + 
\nonumber \\ 
1728\, \sqrt{105}\, \psi_0\, \psi_4\, \psi_8^3 - 
612\, \sqrt{70}\, \psi_4^2\, \psi_8^3 - 10080\, \psi_0\, \psi_8^4 - 
2520\, \sqrt{6}\, \psi_4\, \psi_8^4 + 1260\, \sqrt{70}\, \psi_8^5\, .
\end{eqnarray}

\begin{figure}[htb] \centering
\mbox{  \epsfysize=100mm
         \epsffile{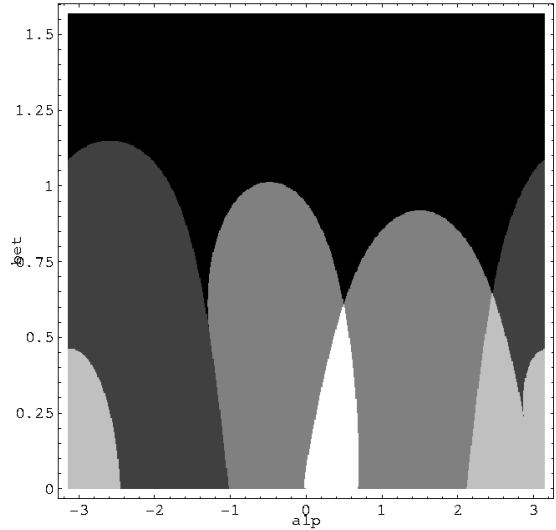}
     }
\caption{Mixture of $H_0,$ $H_4$ and $H_8.$ White, no real positive $\rho$ 
root; black, 4 roots; darkening grays, increasing root number.} 
\end{figure}

It is convenient to take advantage of the free scaling of ${\cal P}$
and normalize the polynomial so that its coefficients lie on a half sphere of
unit radius,
\begin{equation}
\psi_0=\cos \alpha\, \cos \beta,\ \ \psi_4=\sin \alpha\, \cos \beta,\ \ 
\psi_8=\sin \beta,\ \ \ \ -\pi< \alpha \le \pi,\ \ 0< \beta \le \pi/2\, .
\label{sphere2}
\end{equation}
We thus show in Figure 1 the domains where the number of roots increases 
from 0 to 4. The no root domain is the white triangle above $\alpha=\beta=0$ 
and slightly right from this point. In this domain,  $\psi\ (\equiv \varphi)$ is 
self-Fourier and positive.

\subsection{Mixture of four polynomials in the subspace with eigenvalue 1}

Now we add to $\psi$ a component $\psi_{1\!2},$ hence
\begin{eqnarray}
{\cal P} &=& \psi_0 + \frac{\psi_4}{2\, \sqrt{6}}\, (4 \rho^2-12 \rho+3) + 
\frac{\psi_8\, }{24\, \sqrt{70}}
\, (16 \rho^4-224 \rho^3+840 \rho^2-840 \rho+105\,) + 
\nonumber \\
& & \frac{\psi_{1\!2}\, }
{1440\, \sqrt{231}}
\, (64 \rho^6-2112 \rho^5+23760 \rho^4-110880 \rho^3+207900 \rho^2-124740 
\rho+10395\,) 
.
\end{eqnarray}
While borders corresponding to $P_+(0)=0$ obtain easily, the resultant 
${\cal R}$ to be considered for other borders is unwieldy and is skipped
here. Taking advantage of scaling we set,
\begin{eqnarray}
\psi_0 = \cos \alpha\, \cos \beta\, \cos \gamma,\ \  
\psi_4 &=& \sin \alpha\, \cos \beta\, \cos \gamma,\ \   
\psi_8 = \sin \beta \cos \gamma,\ \ 
\psi_{1\!2} = \sin \gamma, 
\nonumber \\  
-\frac{\pi}{2}< \alpha \le \frac{\pi}{2} &,&\ -\pi < \beta \le \pi\, ,\ \ \ 
0< \gamma \le \frac{\pi}{2}\, .
\label{sphere3}
\end{eqnarray}

This choice of spherical coordinates was designed to ensure the positivity of
$\psi_{1\!2},$ obviously, but also a dominance of $\psi_0$ near 
$\alpha\!=\!\beta\!=\!\gamma\!=\!0.$ The dominance is clearly useful for the 
positivity of ${\cal P}.$ Then this $S_3$ sphere can be explored by various 
cuts according to fixed values of $\gamma.$ The results are shown in Figures 
2-5, with 
$\gamma=\pi/10^6,\, \pi/15,\, 2 \pi/15,\, \pi/6,\, \pi/5,\, 7\pi/30,\,
4\pi/15,\, 3\pi/10\, ,$ respectively. The color code for the number of 
roots is: 0 root, red; 1, yellow; 2, yellowish green; 3, bluish green; 4, 
blue; 5, dark purple; 6, pink (in the uncolored edition, the no root domain, 
if it exists, is that dark, small or tiny triangle slightly right of the map 
center). The red domain shrinks at first very slowly when $\gamma$ increases, 
then faster when $\gamma \simeq \pi/6.$ Beyond such an order of magnitude for 
$\gamma,$ there is no red domain and the map becomes invaded by bigger and 
bigger pink patches, representing the dominance of the $6$ positive, real 
roots of $H_{1\!2}.$ 

\begin{figure}[htb] \centering
\mbox{  \epsfysize=100mm
         \epsffile{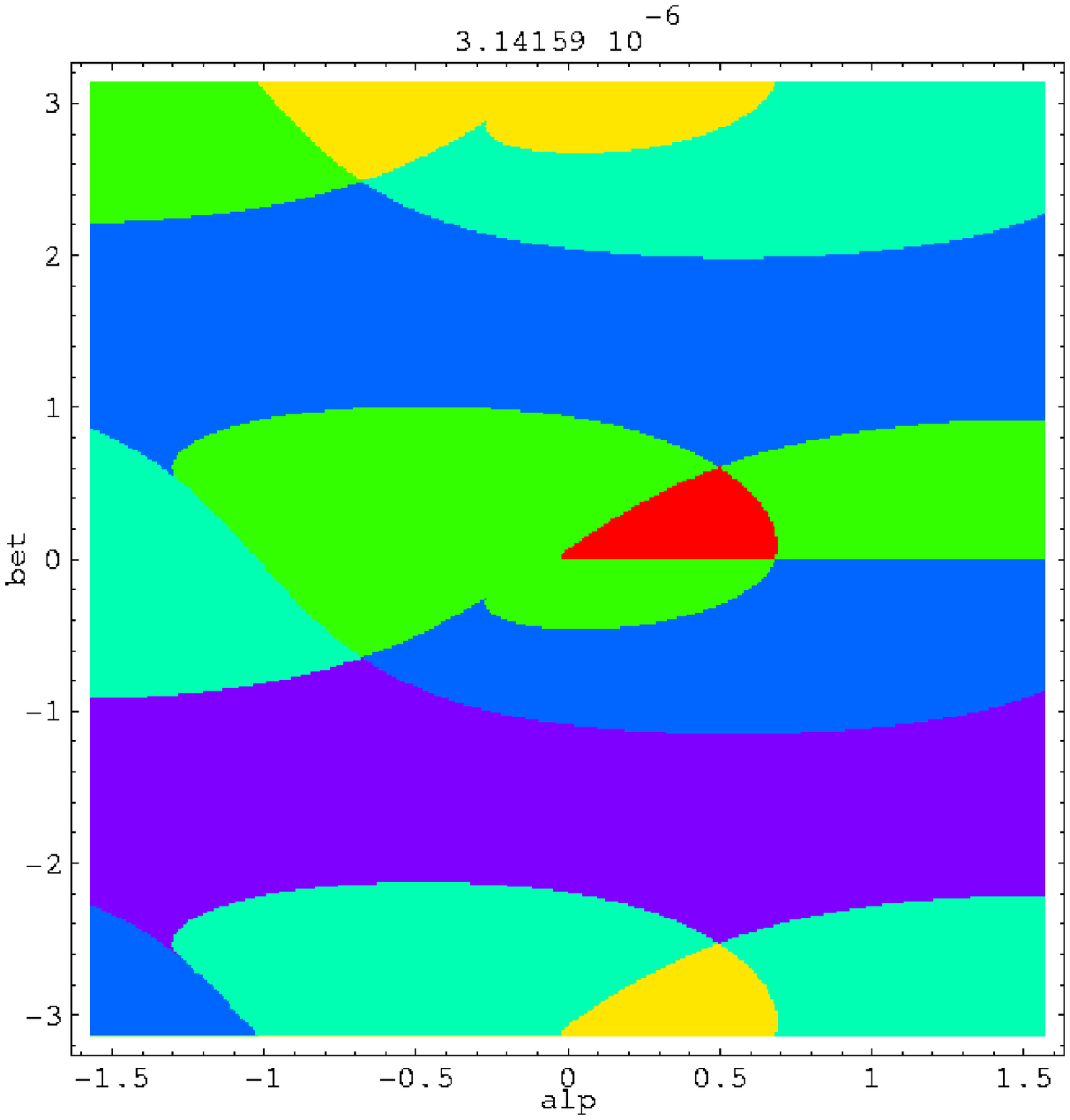}
     }
\mbox{  \epsfysize=100mm
         \epsffile{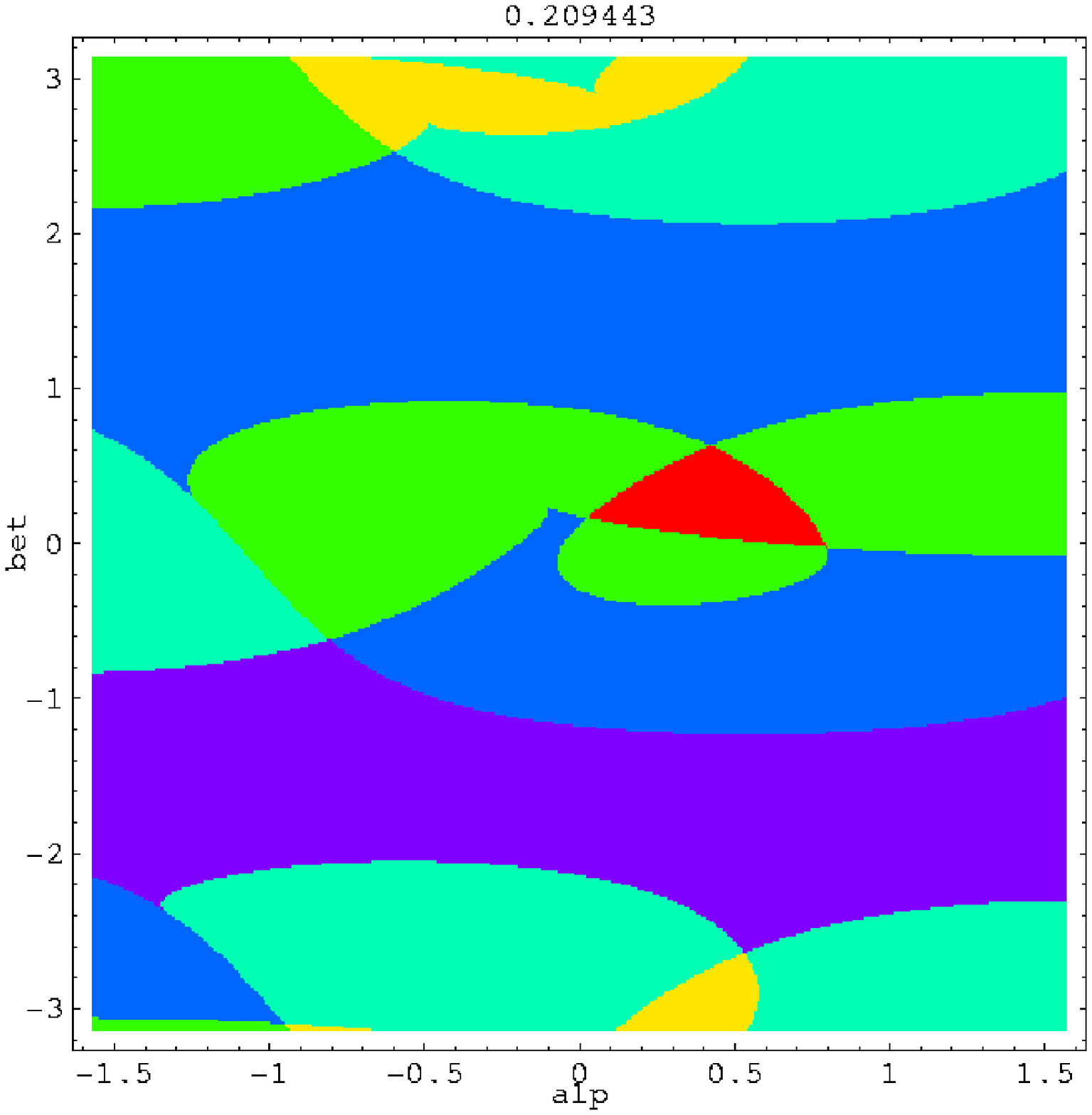}
     }
\caption{Mixture of $H_0,$ $H_4,$ $H_8$ and $H_{1\!2}.$ Root number maps.
Color code: 0 root, red; 1, yellow; 2, yellowish green; 3, bluish green; 
4, blue; 5, dark purple; 6, pink. Left: cut of the parameter sphere $S_3$ 
when $\gamma=\pi/10^6.$ Right: cut for $\gamma=\pi/15$ (uncolored edition: 
the $0$ root domain is the dark triangle-like domain near the center).} 
\end{figure}

\begin{figure}[htb] \centering
\mbox{  \epsfysize=100mm
         \epsffile{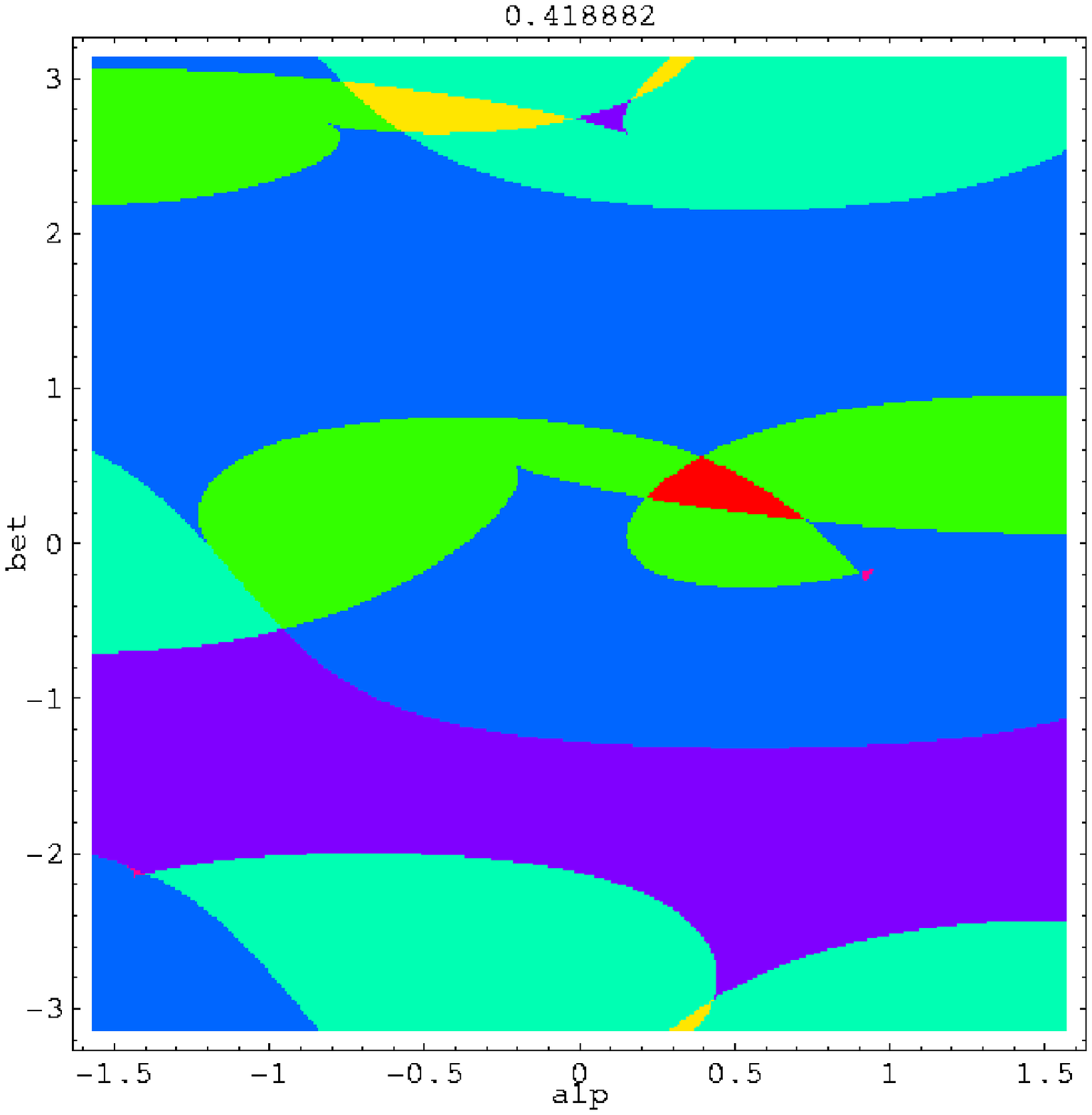}
     }
\mbox{  \epsfysize=100mm
         \epsffile{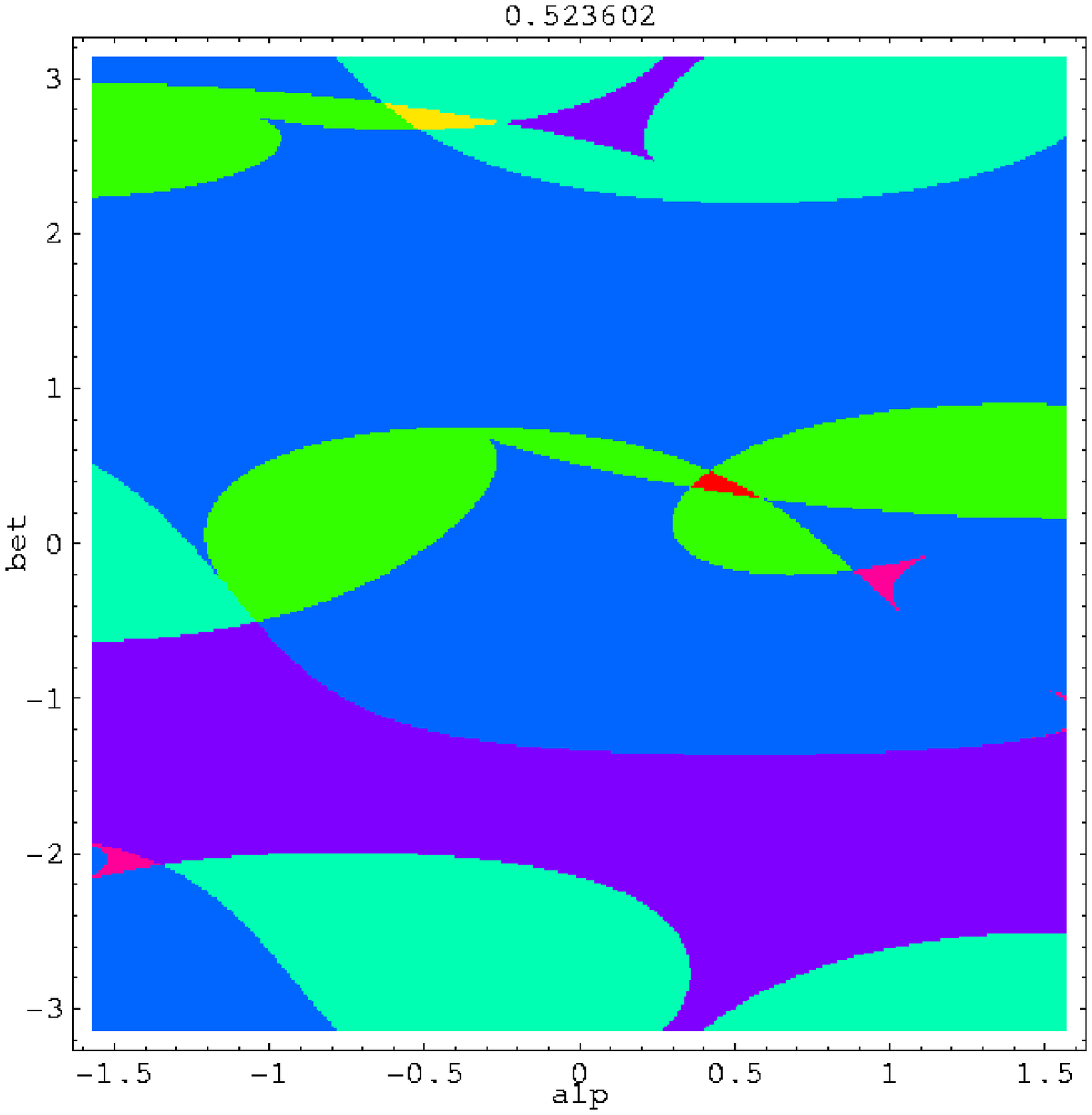}
     }
\caption{Left: Same as Fig. 2, with $\gamma=2\pi/15.$ Right: $\gamma=\pi/6.$
See how the no root triangle shrinks.} 
\end{figure}

\begin{figure}[htb] \centering
\mbox{  \epsfysize=100mm
         \epsffile{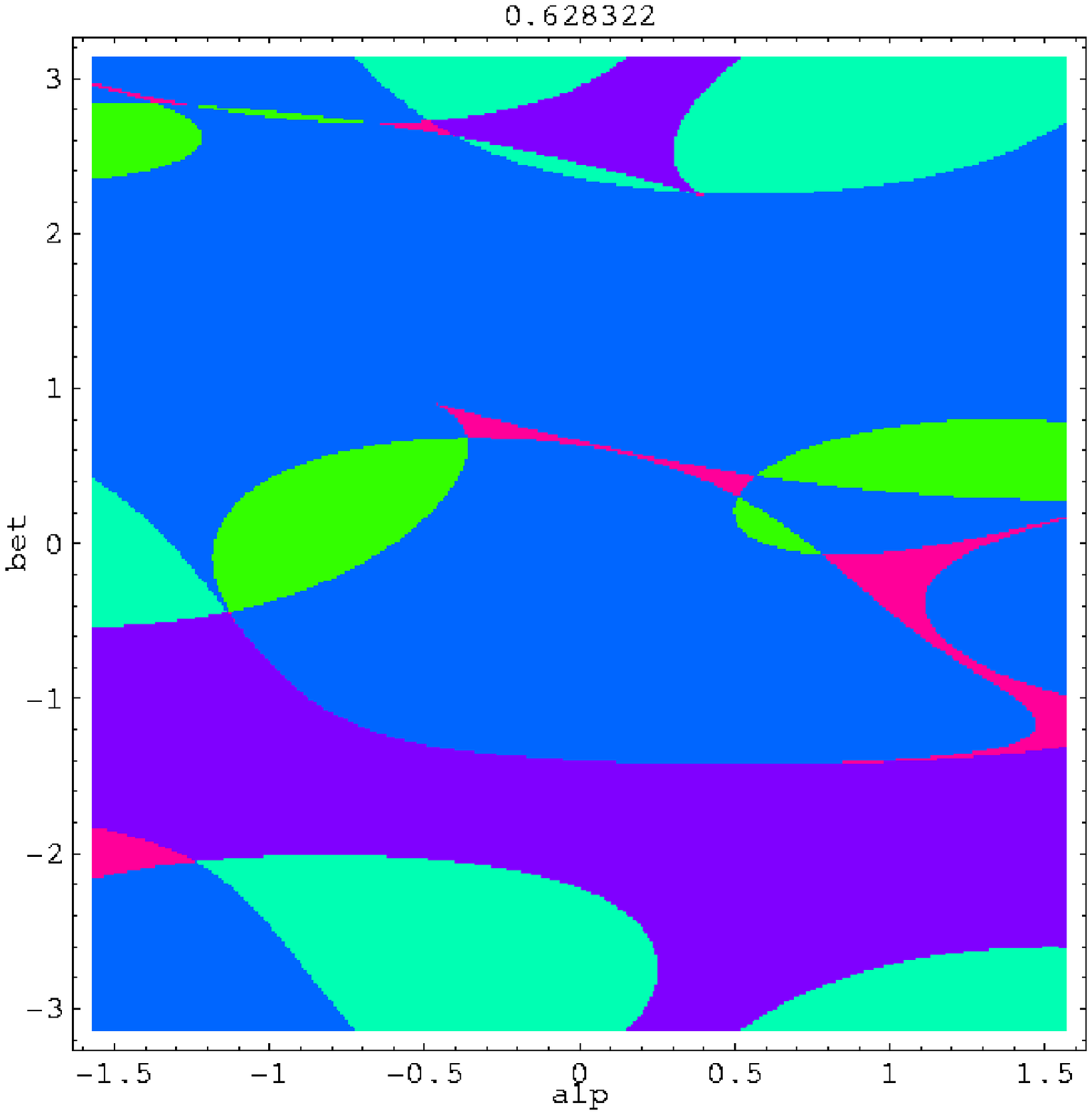}
     }
\mbox{  \epsfysize=100mm
         \epsffile{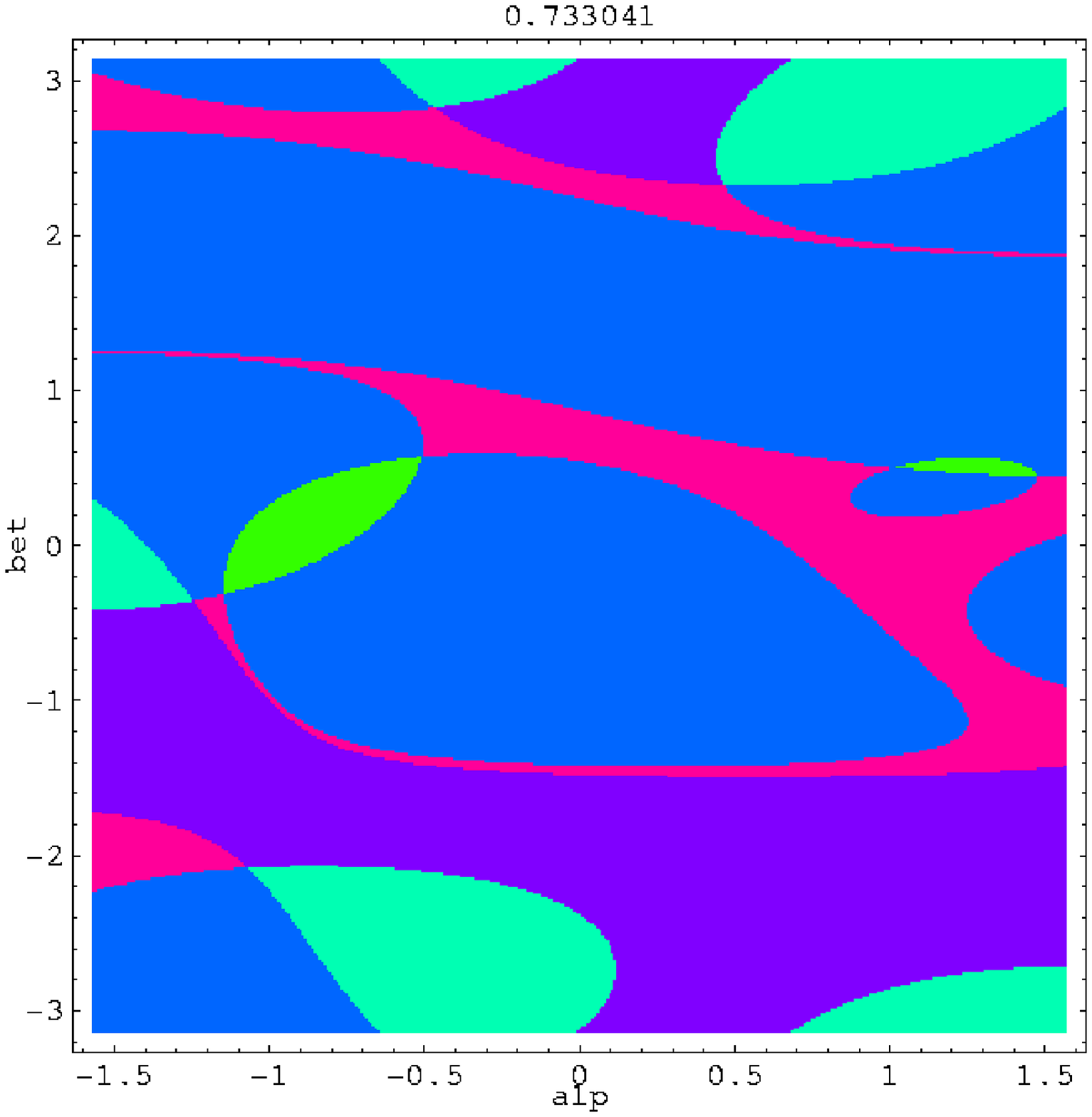}
     }
\caption{Left: $\gamma=\pi/5;$ Right: $\gamma=7\pi/30.$ Absence of no root 
domain.} 
\end{figure}

\begin{figure}[htb] \centering
\mbox{  \epsfysize=100mm
         \epsffile{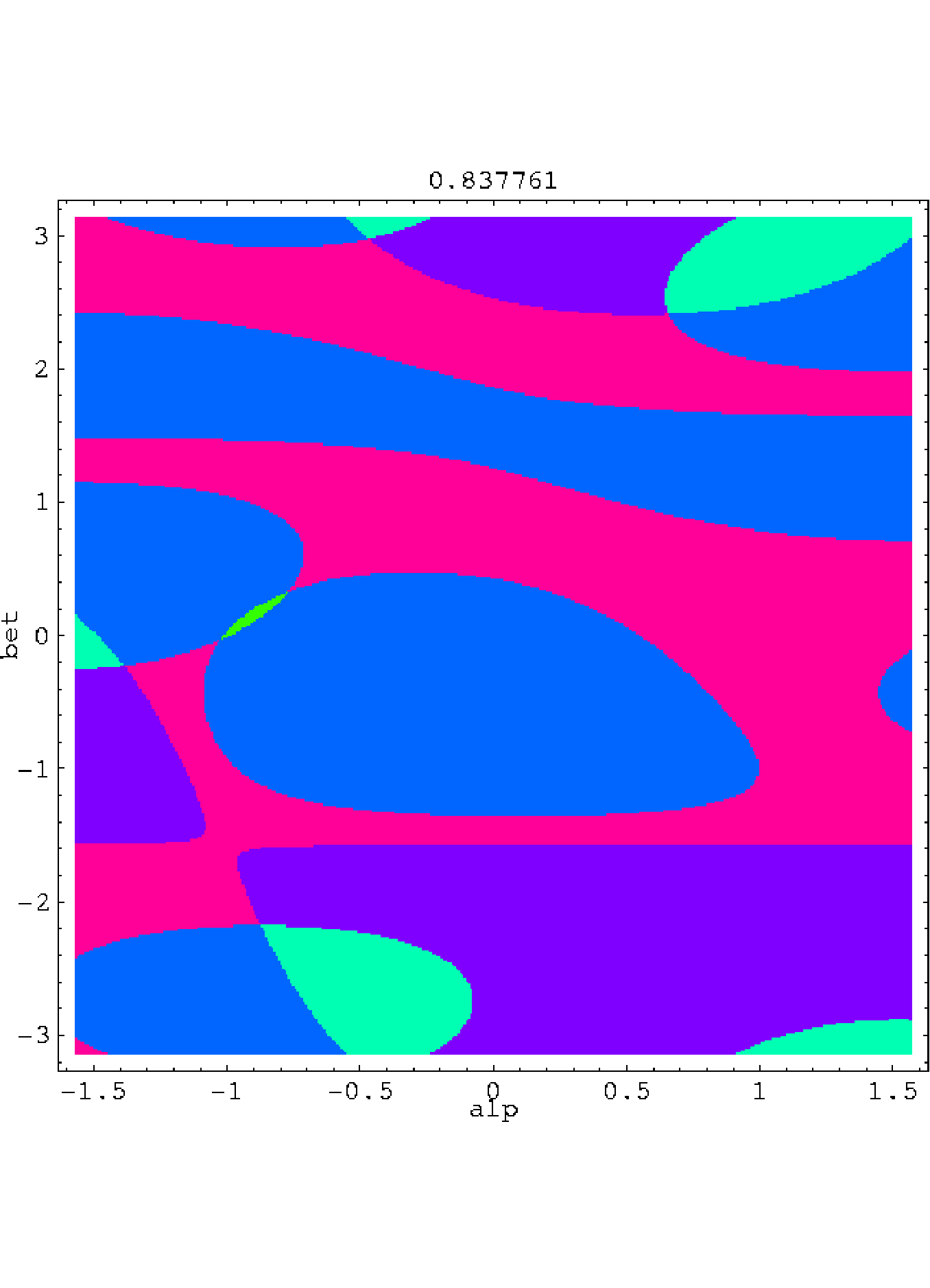}
     }
\mbox{  \epsfysize=100mm
         \epsffile{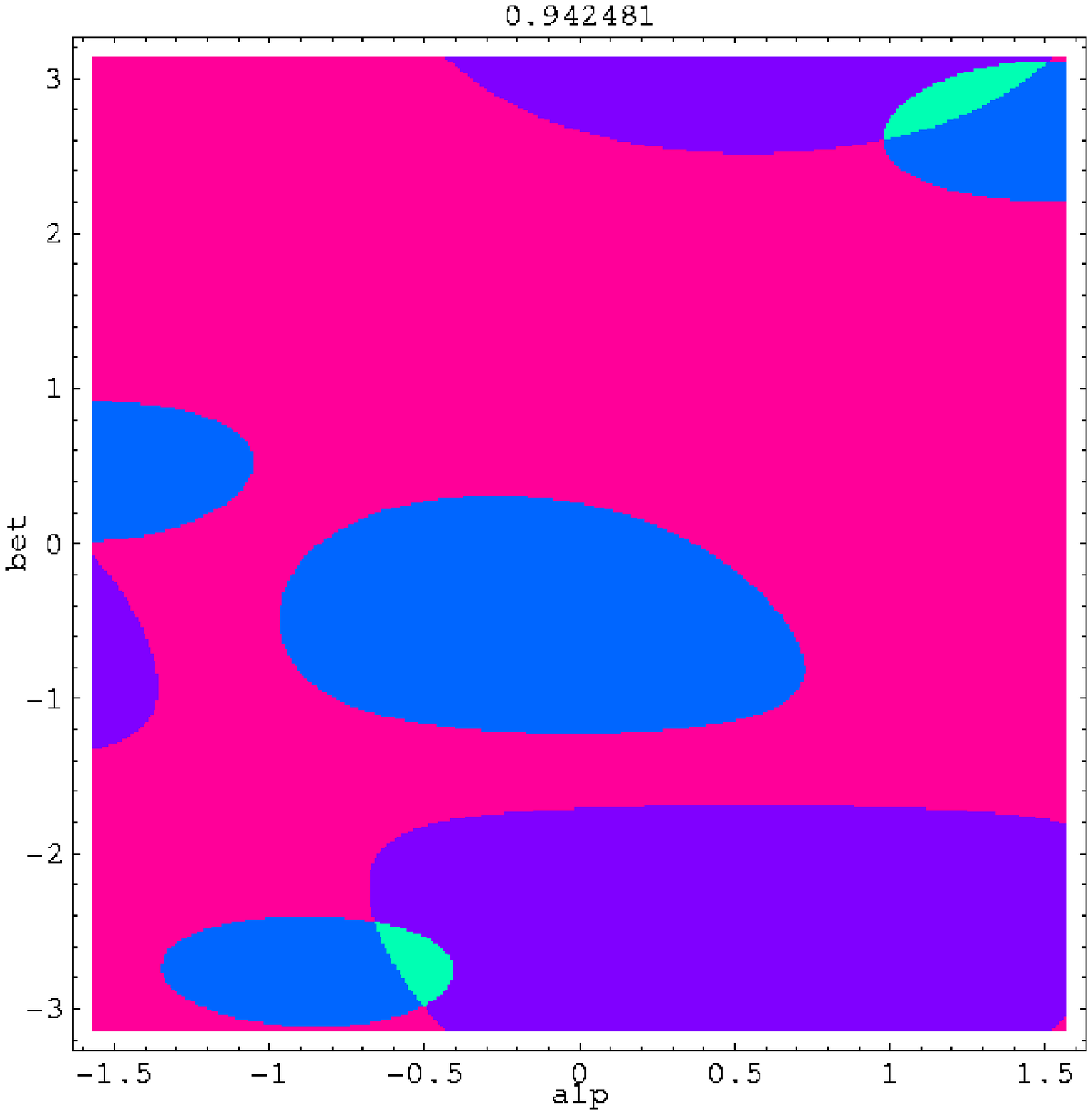}
     }
\caption{Left: $\gamma=4\pi/15;$ Right: $\gamma=3\pi/10.$ Progressive 
dominance of $H_{1\!2}$ with 6 $\rho$ roots. } 
\end{figure}

\subsection{Two polynomials from subspace ``$1$'' mixed with one polynomial 
from subspace ``$-1$''}

If we consider a mixture of $H_0,$ $H_2$ and $H_4,$ the FT connects the two
polynomials
\begin{equation}
{\cal P}(\rho) = \psi_0 + \psi_2\, \frac{2 \rho -1}{\sqrt{2}} 
+ \psi_4\, \frac{4 \rho^2-12 \rho+3}{2\, \sqrt{6}}\, ,
\ \ \ \ 
{\cal Q}(\sigma) = \psi_0 - \psi_2\, \frac{2 \sigma -1}{\sqrt{2}} 
+ \psi_4\, \frac{4 \sigma^2-12 \sigma+3}{2\, \sqrt{6}}\, .
\end{equation}
We study the positivity of each polynomial separately, then of both. 
Notice that the parametrization,
\begin{equation}
\psi_0=\cos \alpha\, \cos \beta,\ \ \psi_2=\sin \alpha\, \cos \beta,\ \ 
\psi_4=\sin \beta,\ \ \ \ -\pi< \alpha \le \pi,\ \ 0< \beta \le \pi/2\, ,
\end{equation}
reverses only the sign of $\psi_2$ if $\alpha$ becomes $-\alpha$. This 
parity operation is seen in  Figure 6, the white domains 
of which correspond to the positivity of ${\cal P}$ and ${\cal Q},$ 
respectively. The domain of simultaneous positivity for both is the 
white intersection domain in the left part of Figure 7, with 
the expected symmetry.

It is actually easy here to analyze analytically the resultants of 
interest for ${\cal P}$ and ${\cal Q},$
\begin{equation}
{\cal R} \propto 
\sqrt{6}\, \psi_2^2 - 4\, \psi_0\, \psi_4\ \  \mp \ 4\, \sqrt{2}\, \psi_2\,
\psi_4 + 2\, \sqrt{6}\, \psi_4^2,
\end{equation}
together with signatures for the signs of roots, such as ${\cal P}(0),$ etc.
This can be done also in the ``spherical representation''. The white domains 
of Figs. 6, 7 are recovered.

\begin{figure}[htb] \centering
\mbox{  \epsfysize=100mm
         \epsffile{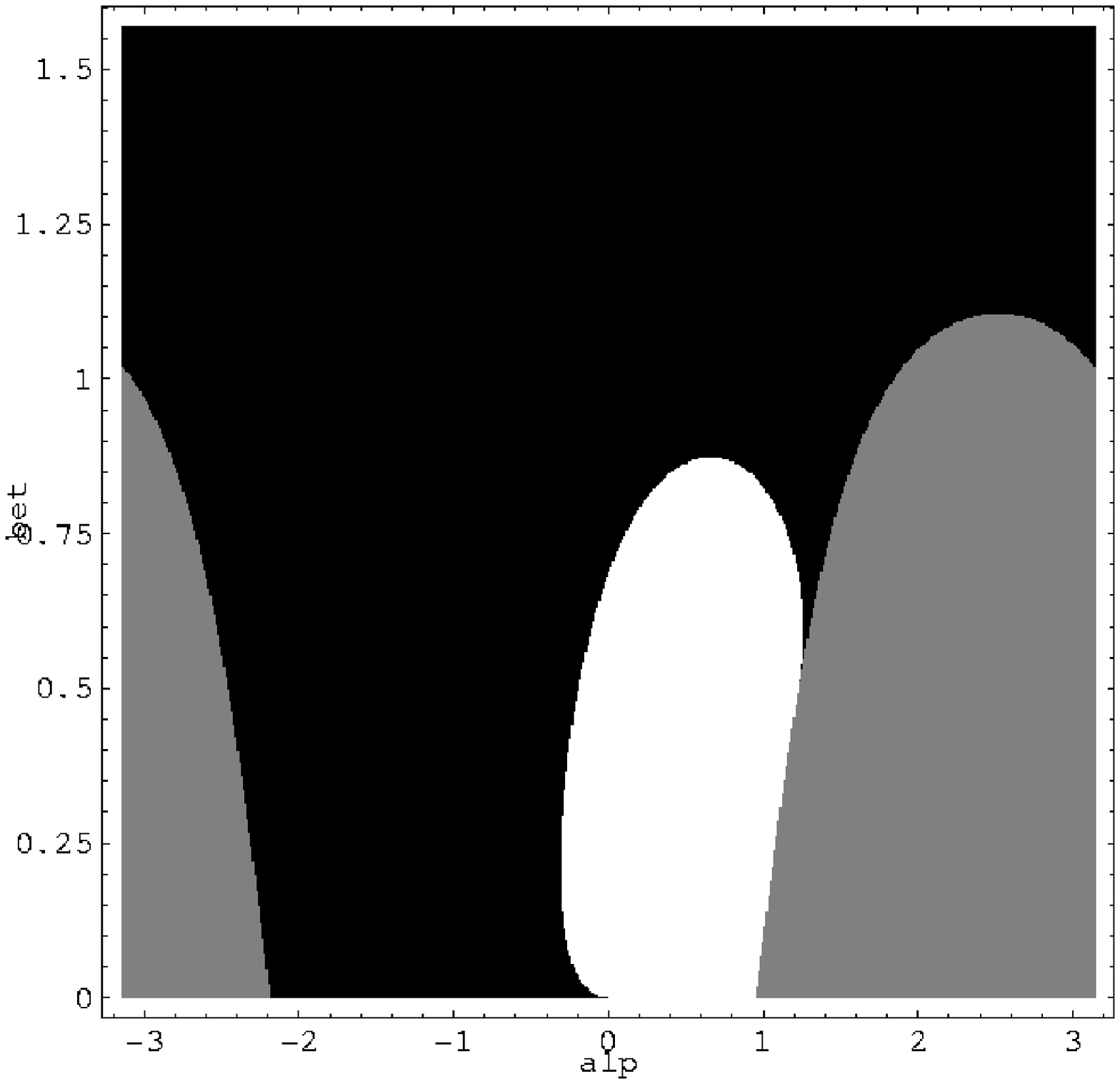}
     }
\mbox{  \epsfysize=100mm
         \epsffile{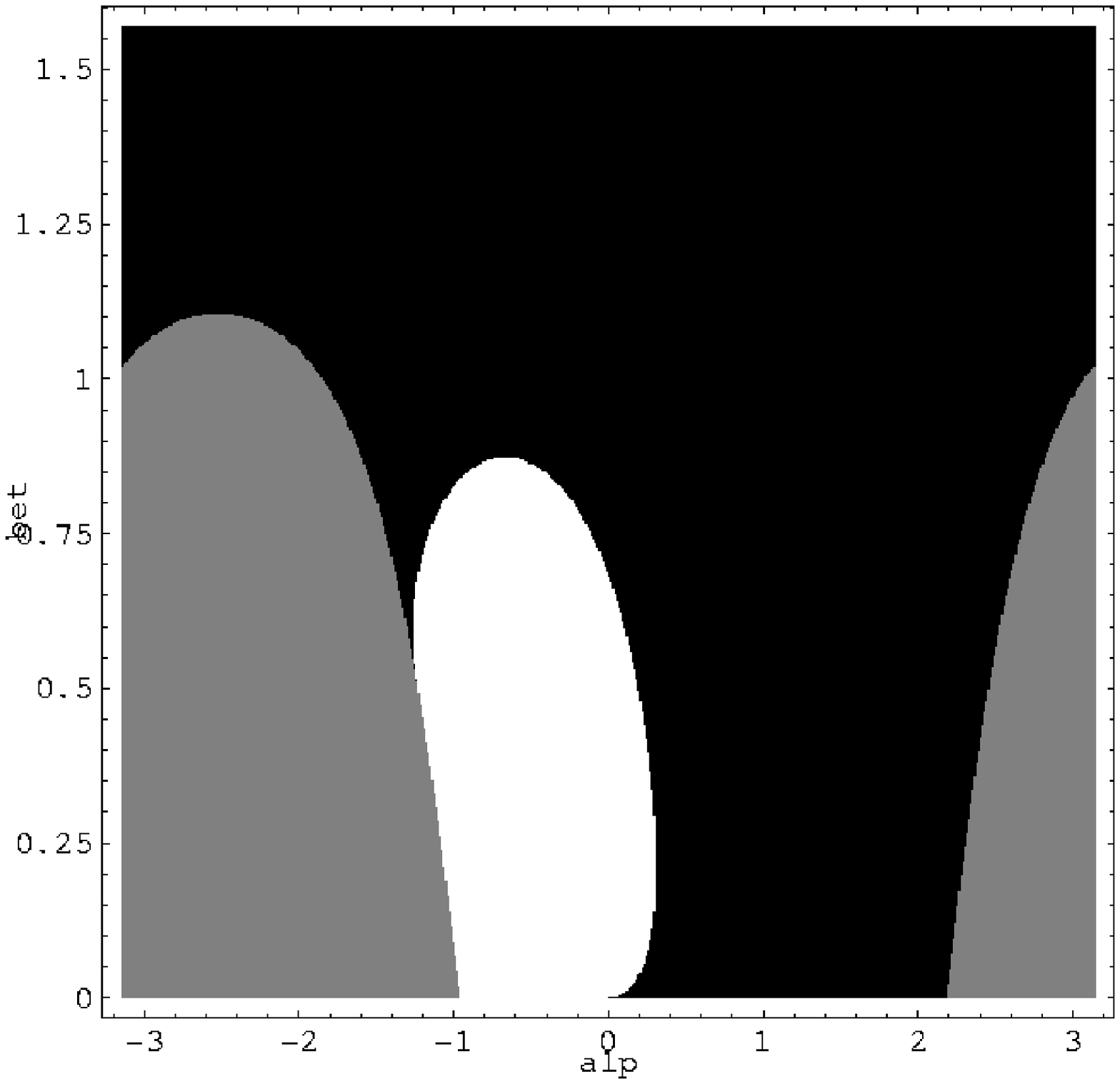}
     }
\caption{Mixture of $H_0,$ $H_2$ and $H_4.$ White domain, 0 root. Grey, $1$ 
root; black, $2$ roots. Left: results for ${\cal P}$; Right:  ${\cal Q}.$ 
} 
\end{figure}

\begin{figure}[htb] \centering
\mbox{  \epsfysize=100mm
         \epsffile{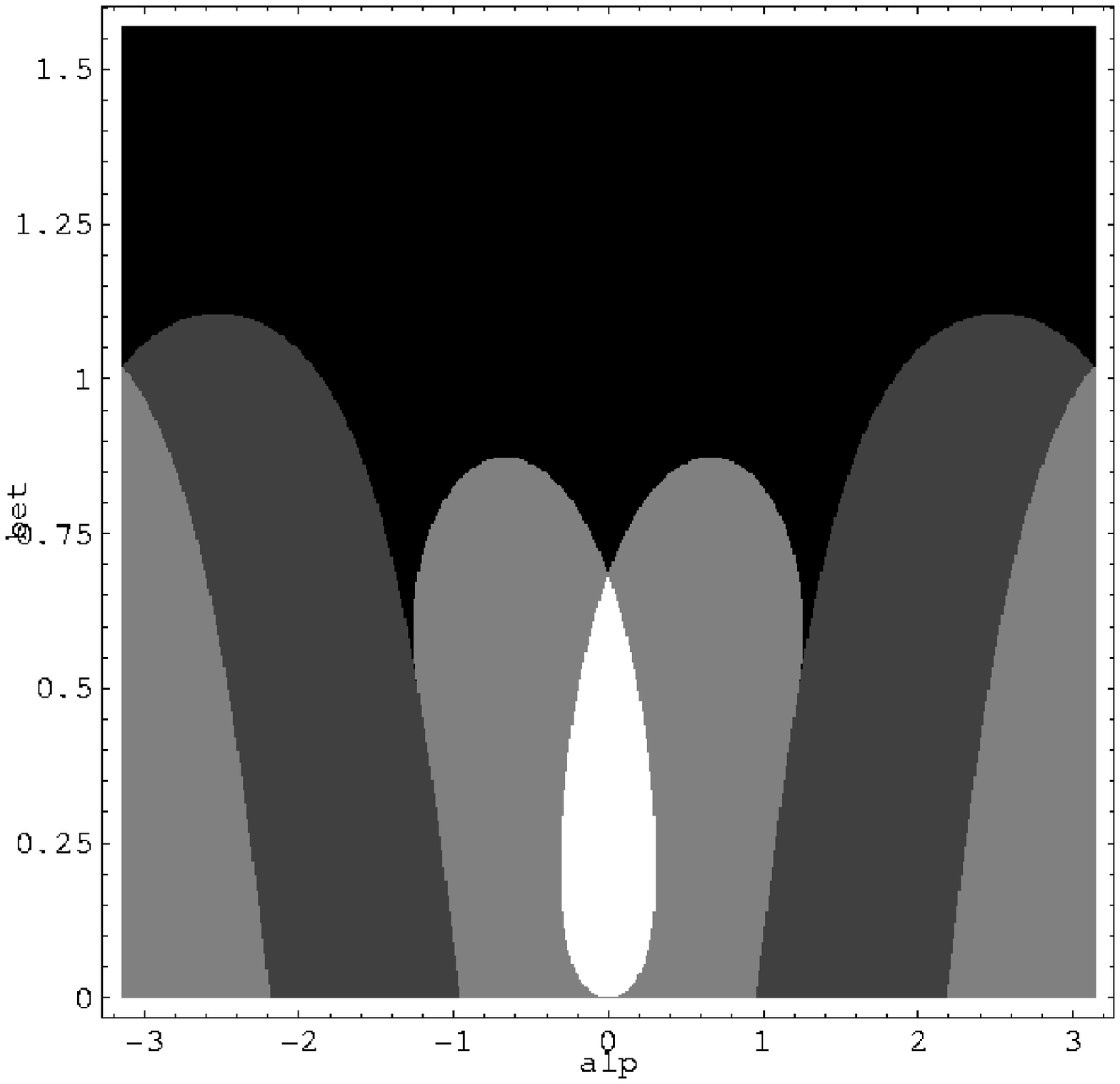}
     }
\mbox{  \epsfysize=100mm
         \epsffile{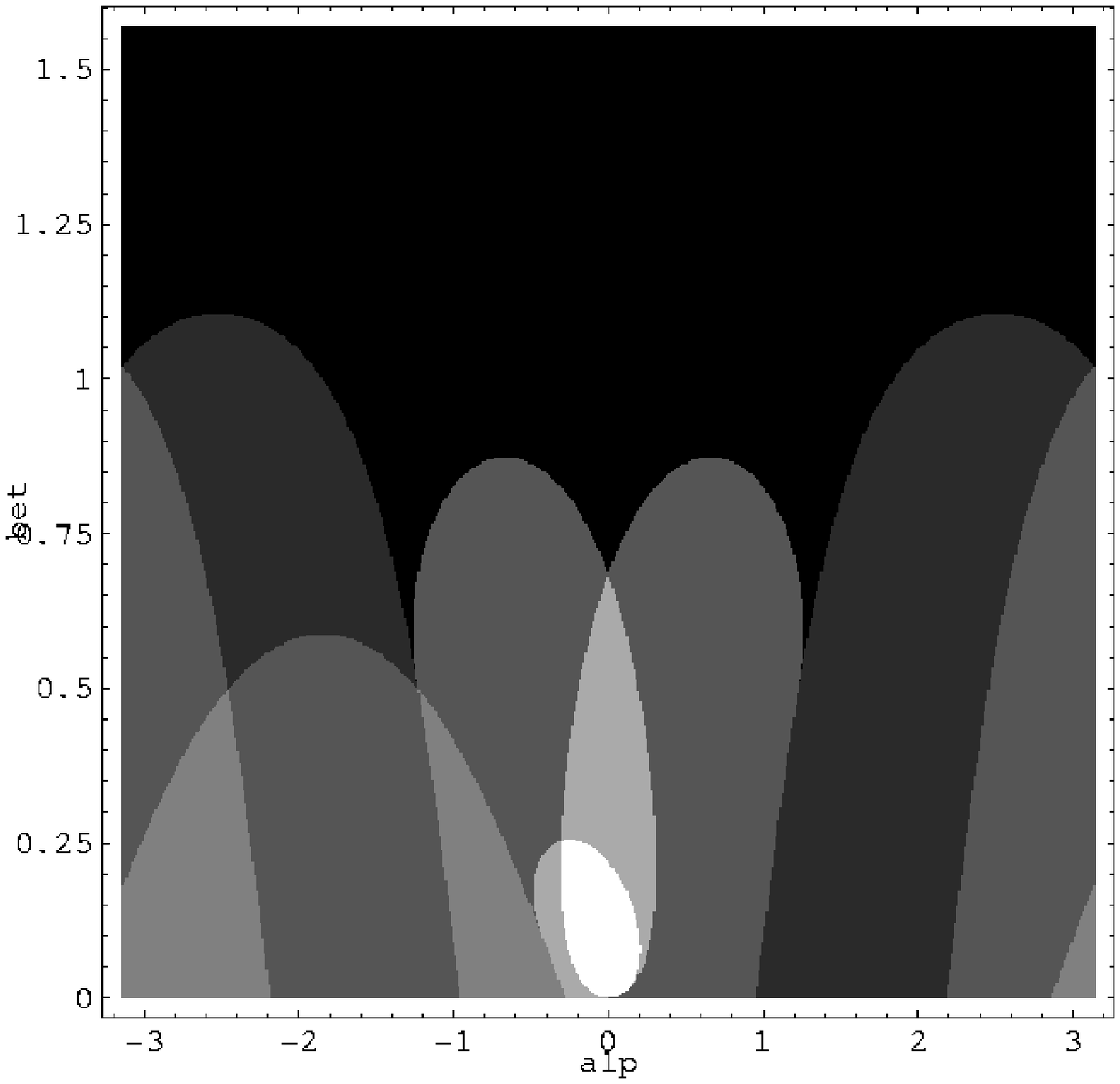}
     }
\caption{Positivity domains for the parity mixed case. Left: for both 
${\cal P}$ and ${\cal Q}.$ Right: ${\cal P},$ ${\cal Q}$ and monotony of 
$\varphi.$}
\end{figure}

\subsection{Positivity with monotony}

For some problems \cite{Kovchegov}, it may be useful to request either 
$\psi$ and/or $\varphi$ to be monotonous functions in an interval such as 
$[ 0,\infty [ .$ We illustrate this in the case of an $H_0,H_2,H_4$ mixture, 
with the additional constraint,
\be
\frac{d \varphi}{d \sigma} \propto -12\, \cos \alpha\, \cos \beta + 6\, 
\sqrt 2\, \cos \beta\, \sin \alpha\, (2 \sigma - 5) - \sin \beta\, \sqrt 6\,
(4 \sigma^2 - 28 \sigma + 27) <0\, . 
\ee
The result appears in the right part of Fig. 7. The white domain, 
corresponding to such three simultaneous conditions of positivity and 
monotonicity, is a severe restriction of the white domain seen in the 
left part of Fig. 7.

\subsection{Positivity from convexity}

A practical and sufficient, but not necessary condition for the positivity 
of $\varphi$ is the convexity of $\psi$ \cite{lafforgue}. Indeed,
$\varphi\sim \int_0^{\infty}dr\, [1-\cos(sr)]/s^2\, d^2\psi/dr^2 >0.$
We illustrate this convexity condition for a mixture of 
$H_0,H_4,H_8.$ It is clear that the presence of 
$\exp\!\left(-r^2/2\right)$ in front of a finite order polynomial, 
with the even parity of $\psi$ and its derivability, are contradictory with 
``convexity everywhere''; a smooth, round maximum must occur at the origin. 
We thus study partial convexity conditions of the kind, ``convexity between 
$r_c$ and $+\infty$''. A reasonable choice for the order of magnitude of 
$r_c$ is the position $r_c=1$ of the inflexion point of 
$\exp\!\left(-r^2/2\right).$ The second derivative $d^2 \psi/dr^2$ belongs 
to the same algebra. We adjusted its Sturm criterion to various values of 
$r_c.$ Small domains only are found that ensure zero roots for 
$d^2 \psi / dr^2,$ because most mixtures of 
$H_0,H_4,H_8$ do oscillate. Figures 8 and 9 show, in white again, with the 
parametrization by Eqs. (\ref{sphere2}), the survivor domain obtained if 
$r_c=1,$ right part Fig. 8, then $r_c^2=2/3,$ left part Fig. 9, and 
$r_c^2=2/5,$ right part Fig. 9, respectively. The domain shrinks in a smooth 
way when $r_c^2$ decreases from $1$ to $0.4$ and disappears if 
$r_c^2 < \sim 0.4.$ It does not increase much when  $r_c \ge {\cal O} (1).$ 
The left part of Fig. 8, a zoom of the left part of Fig. 6, shows that such 
partial convexity domains are already included in the positivity domain of 
$\psi.$

\begin{figure}[htb] \centering
\mbox{  \epsfysize=100mm
         \epsffile{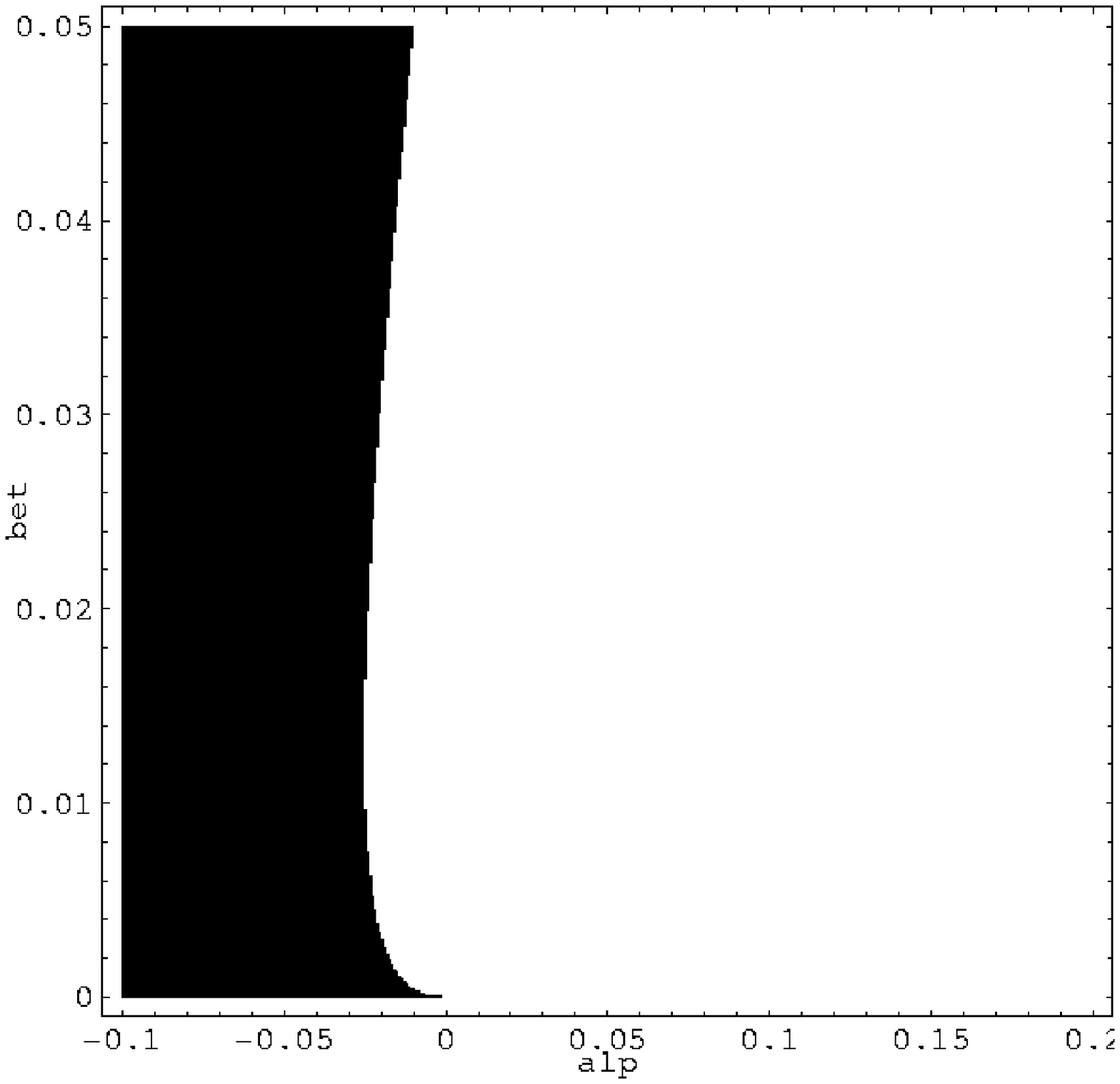}
     }
\mbox{  \epsfysize=100mm
         \epsffile{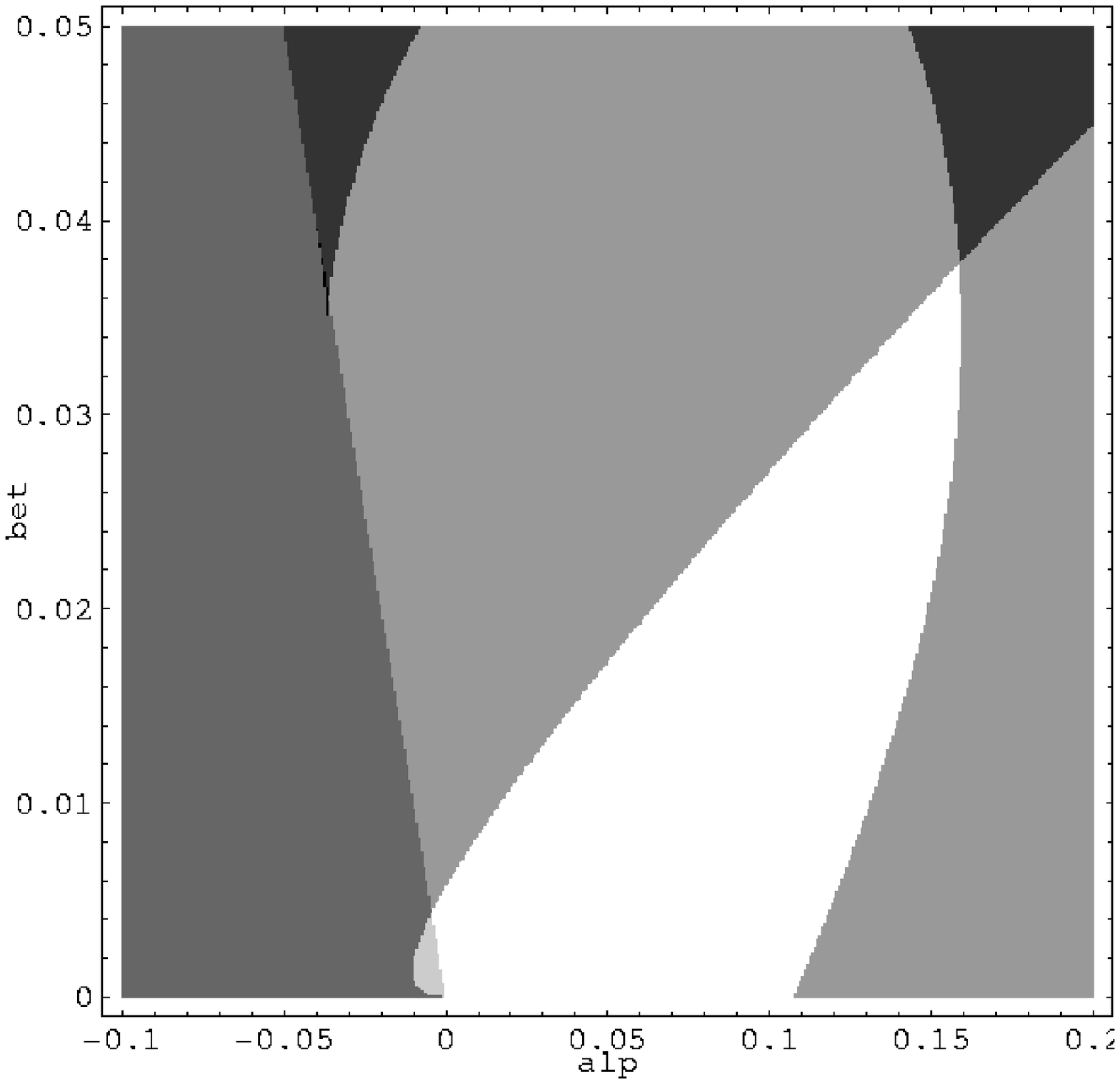}
     }
\caption{$H_0,H_4,H_8$ mixture. Left: zoom of the positivity domain. 
Right: convexity domain for $r > r_c=1.$}
\end{figure}

\begin{figure}[htb] \centering
\mbox{  \epsfysize=100mm
         \epsffile{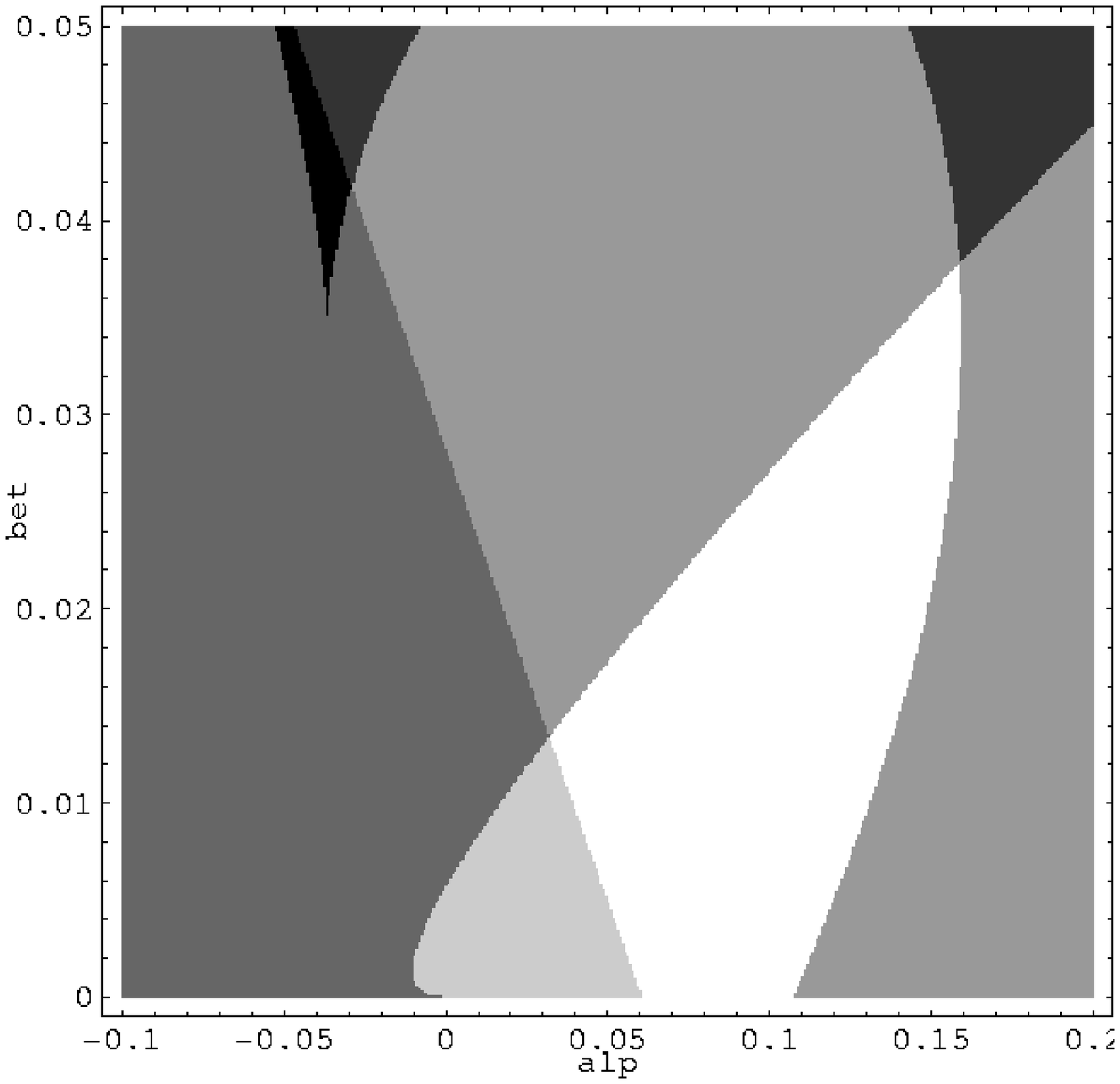}
     }
\mbox{  \epsfysize=100mm
         \epsffile{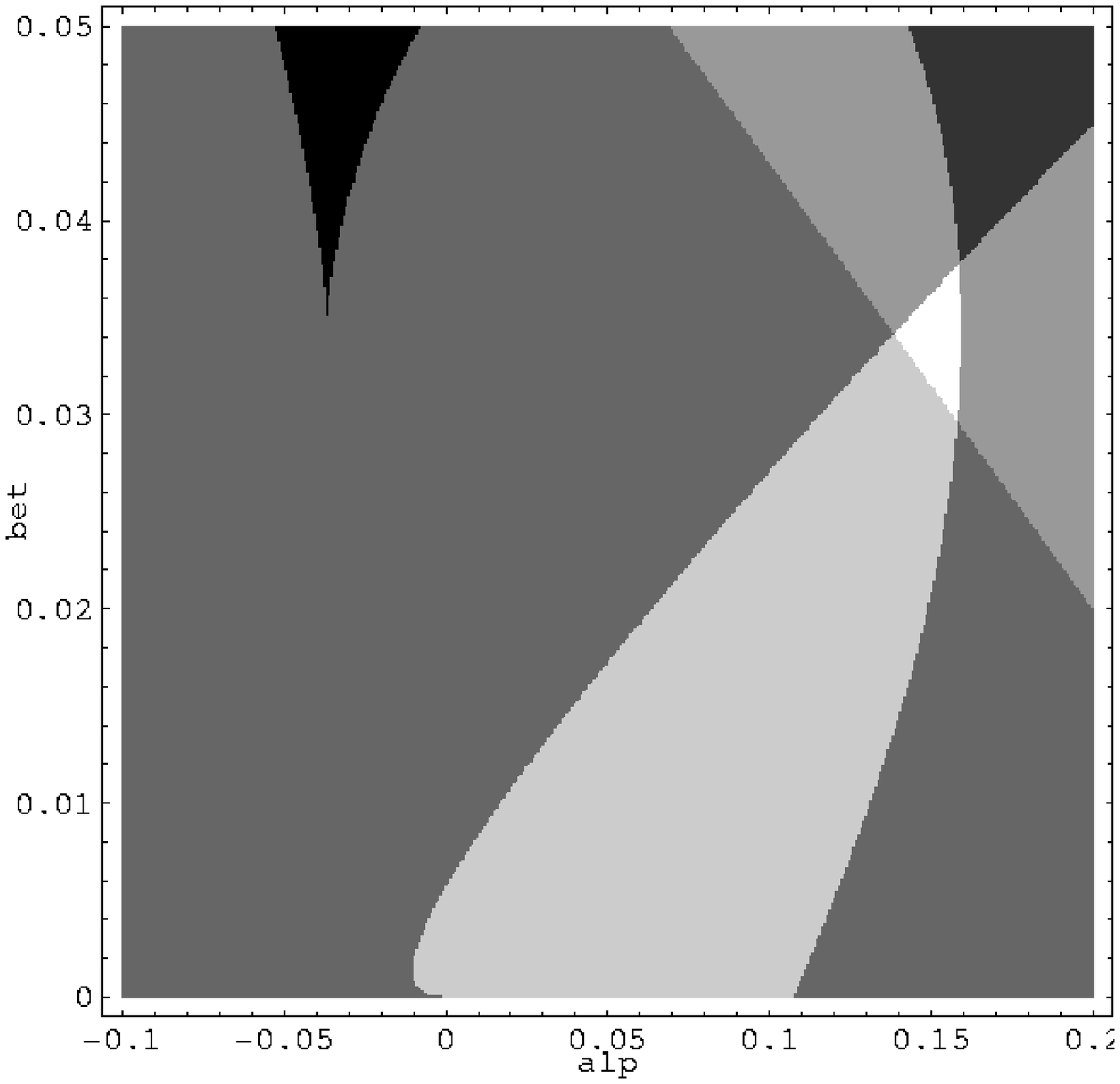}
     }
\caption{Same mixture as Fig. 8. Convexity domains. Left: 
$r_c^2=.67.$ Right: $r_c^2=.4.$}
\end{figure}

\section{Positivity for the 2-dimensional Fourier transform}

The Fourier-Bessel transform in which we are here interested reads,
\be
\varphi(s)=\int_0^{\infty} (r\, dr)\, J_0(s\, r)\, \psi(r)\, .
\ee
For the 2-d radial space, a complete basis of
states results from substituting $r^2$ for $r$ into Laguerre polynomials,
\be
\int_0^{\infty} (r\, dr)\ 2\ e^{-r^2}\, L_m(r^2)\, L_n(r^2) = \delta_{mn}\, .
\label{Lagnorm2}
\ee
For the sake of clarity, we list here the first four such normalized, 
``2-d radial'' states,
\be
\{v_0,\, v_1,\, v_2,\, v_3\}\ =\ \sqrt 2\ e^{-\frac{1}{2}r^2}\, \left\{  
1,\, r^2-1,\,  \frac{r^4-4r^2+2}{2},\,  \frac{r^6-9r^4+18r^2-6}{6}  
\right\}\, .
\ee
One can verify that the states $v_n$ make eigenstates of the Fourier-Bessel 
transform,
\begin{equation}
\int_0^{\infty}(r\, dr)\, J_0(s\, r)\, v_n(r) = (-1)^n\, v_n(s)\, .
\label{transpa2}
\end{equation}
Positivity conditions can again be implemented with the Sturm criterion. 
For instance a mixture of $v_0, v_2,$ from that subspace with eigenvalue 
$1,$ and $v_1,$ from that with eigenvalue $-1,$ defines the following two 
reciprocal partners,
\begin{equation}
{\cal P}(\rho) = 
\psi_0 + \psi_1\, (\rho -1)   + \psi_2\, \frac{\rho^2  -4 \rho  +2}{2}\, ,
\ \ \ \ 
{\cal Q}(\sigma) = 
\psi_0 - \psi_1\, (\sigma -1) + \psi_2\, \frac{\sigma^2-4 \sigma+2}{2}\, .
\end{equation}

\begin{figure}[htb] \centering
\mbox{  \epsfysize=100mm
         \epsffile{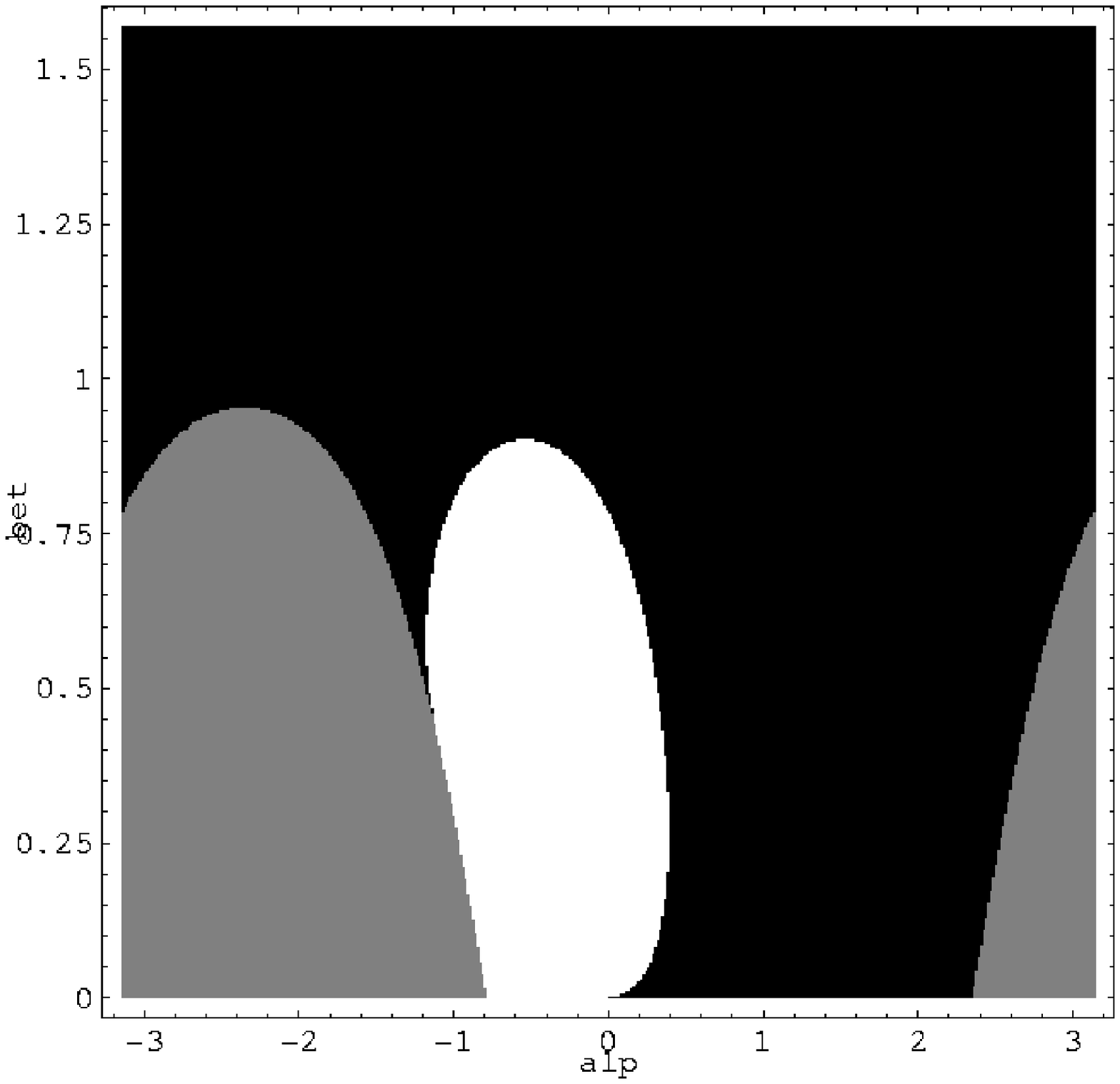}
     }
\mbox{  \epsfysize=100mm
         \epsffile{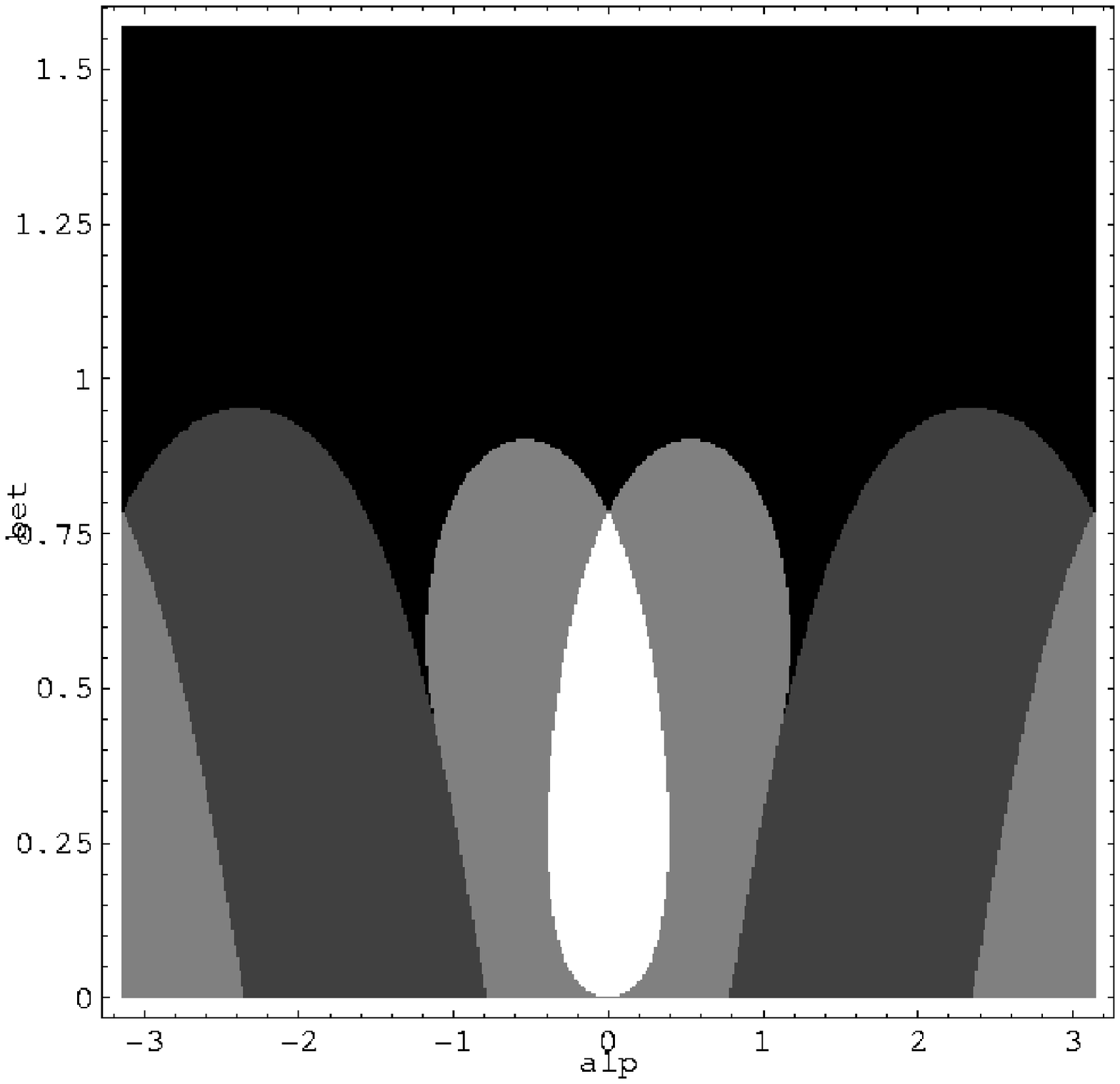}
     }
\caption{Positivity domains for the ``radial 2-d'', parity mixed case. 
Left: $\varphi.$ Right: both $\psi$ and $\varphi.$}
\end{figure}

The left part of Figure 10 shows the positivity domain for $\varphi$ and
its right part shows the joint positivity domain for $\psi$ and $\varphi.$ 
Similarities beween polynomials involved in the present ``mixed parity'' 
case and those of case {\bf C} of the previous Section create topological 
similarities with Fig. 6 (right) and Fig. 7 (left) but numerical details 
do differ.

\begin{figure}[htb] \centering
\mbox{  \epsfysize=100mm
         \epsffile{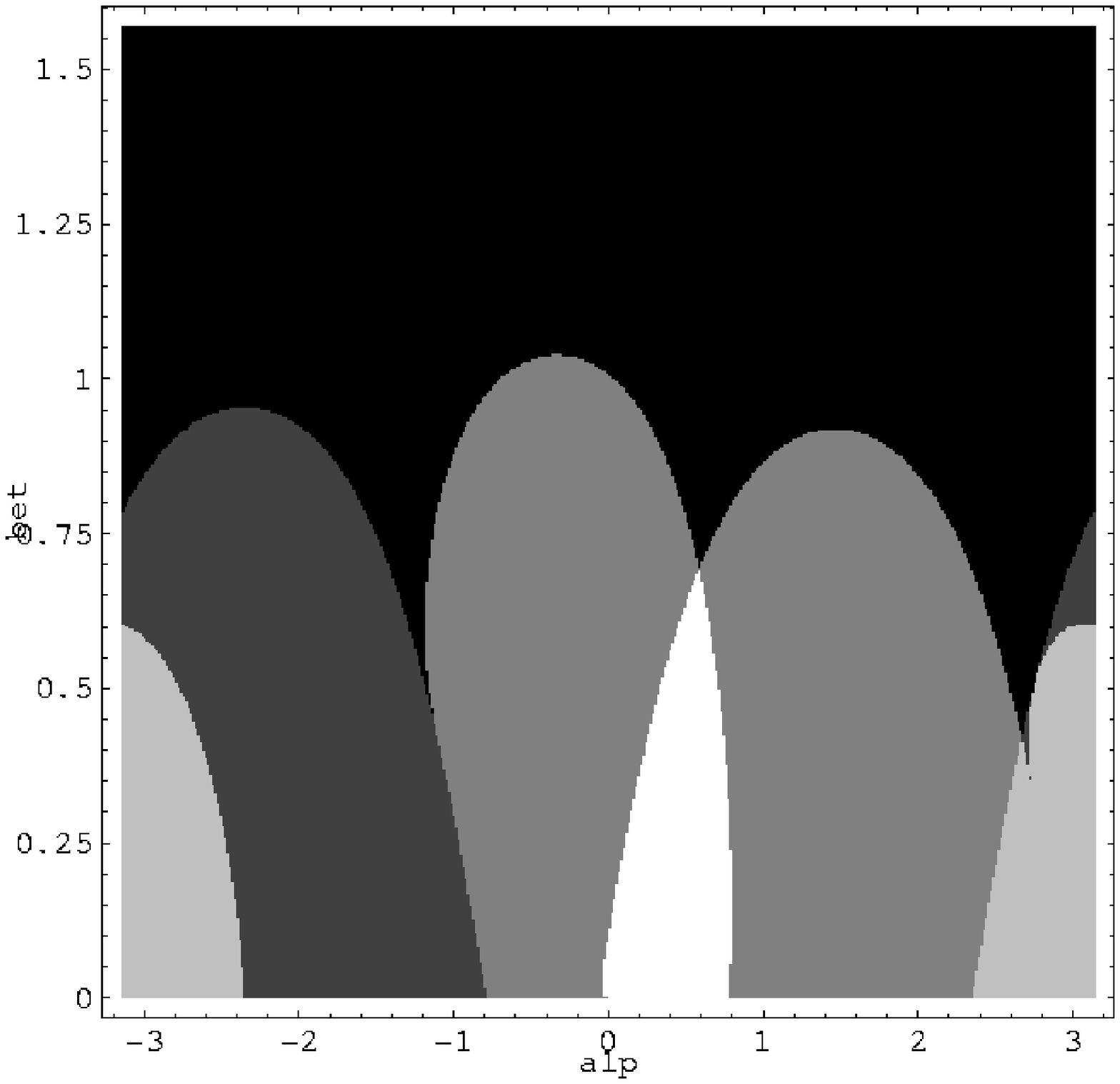}
     }
\mbox{  \epsfysize=100mm
         \epsffile{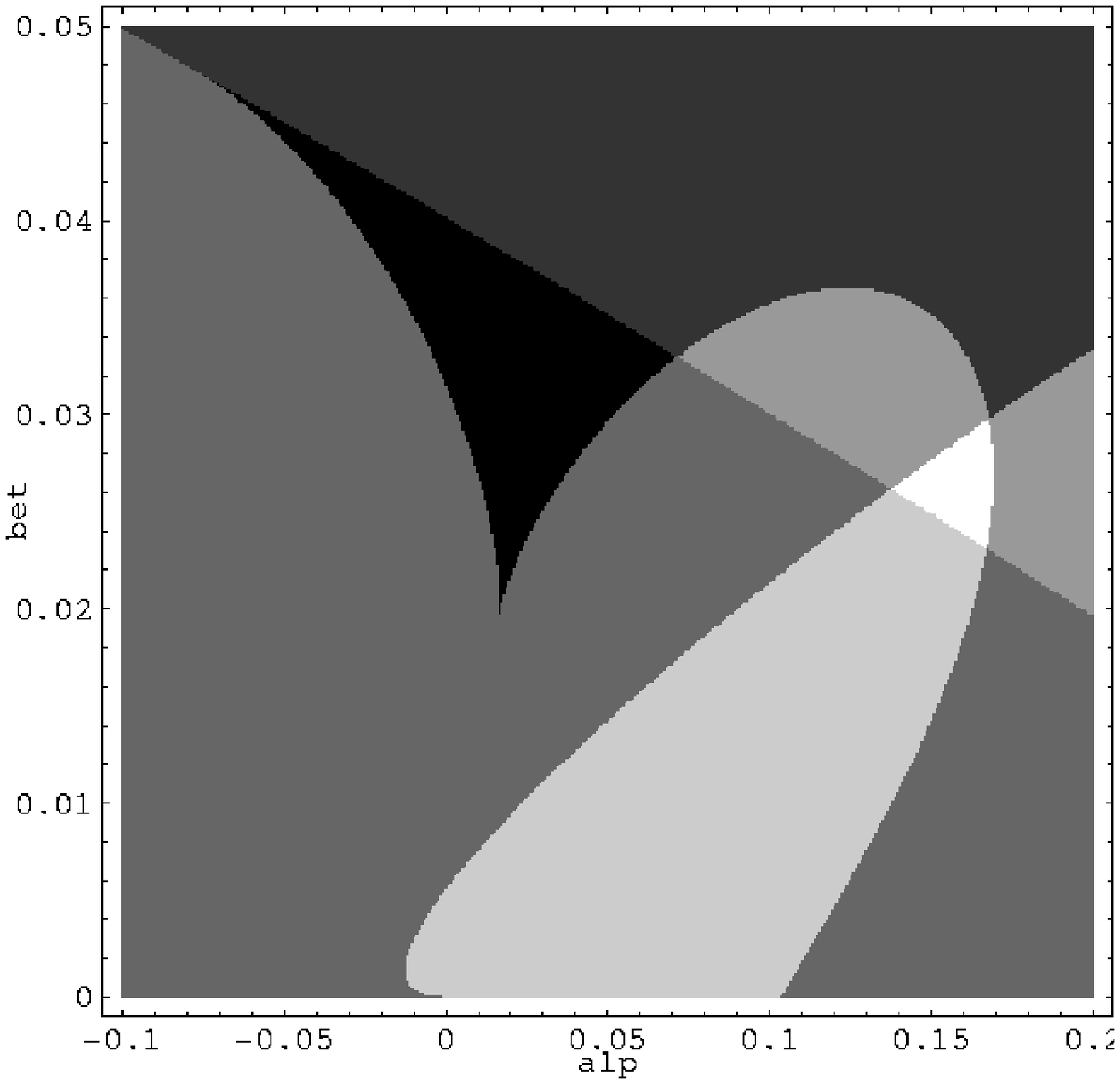}
     }
\caption{Left: Positivity domain for the ``radial 2-d'', $v_0,v_2,v_4$
mixture. Right: third derivative negativity domain if $r_t^2=1.4.$}
\end{figure}

Interestingly enough, a sufficient condition \cite{lafforgue} for the 
positivity of $\varphi$ in this 2-d situation is that the third derivative, 
$d^3 \psi/ dr^3,$ be negative. But, as in the previous Section, 
case {\bf E}, the presence of a truncated number of basis terms in our 
expansion may reduce this negativity condition for $d^3 \psi/ dr^3$ to a 
domain $r \ge r_t$ only. We show in Figure 11 the result for a mixture 
of $v_0,$ $v_2$ and $v_4$ if $r_t^2=1.4,$ a demanding situation resulting 
into the tiny domain in the right part of Fig. 11. The domain belongs to 
the positivity domain seen in the left part of the same Figure.

\section{Discussion and Conclusion}

To summarize our results, we have built a basis of functions 
verifying positivity together with their Fourier transform. The method is 
based upon algebras of Hermite polynomials (for 1-dimensional FT) or 
Laguerre polynomials in the variable $r^2$ (for 2-dimensional radial FT).

The Fourier transform has four eigenvalues $1,i,-1,-i,$ and thus four highly 
degenerate eigensubspaces. Two of such subspaces are compatible with real 
functions remaining real. To span each subspace, we used a basis made of 
Hermite-Fourier (or ``Laguerre-Fourier'') states. The orders (in $r$) of the  
polynomials have to be multiples of $4$ if the eigenvalue is $1,$ and multiple 
of $4$ plus $2$ if the eigenvalue is $-1.$ At the cost of a truncation of such 
bases to a maximum order $N,$ the conditions for positivity, convexity, 
etc. thus reduce to simple manipulations of polynomial coefficients based on 
the Sturm theorem. In each truncation case, one can find suitable domains for 
the parameters which mix the various basis polynomials. Such domains have been 
illustrated by the figures shown by this paper.  

\begin{figure}[htb] \centering
\mbox{  \epsfysize=100mm
         \epsffile{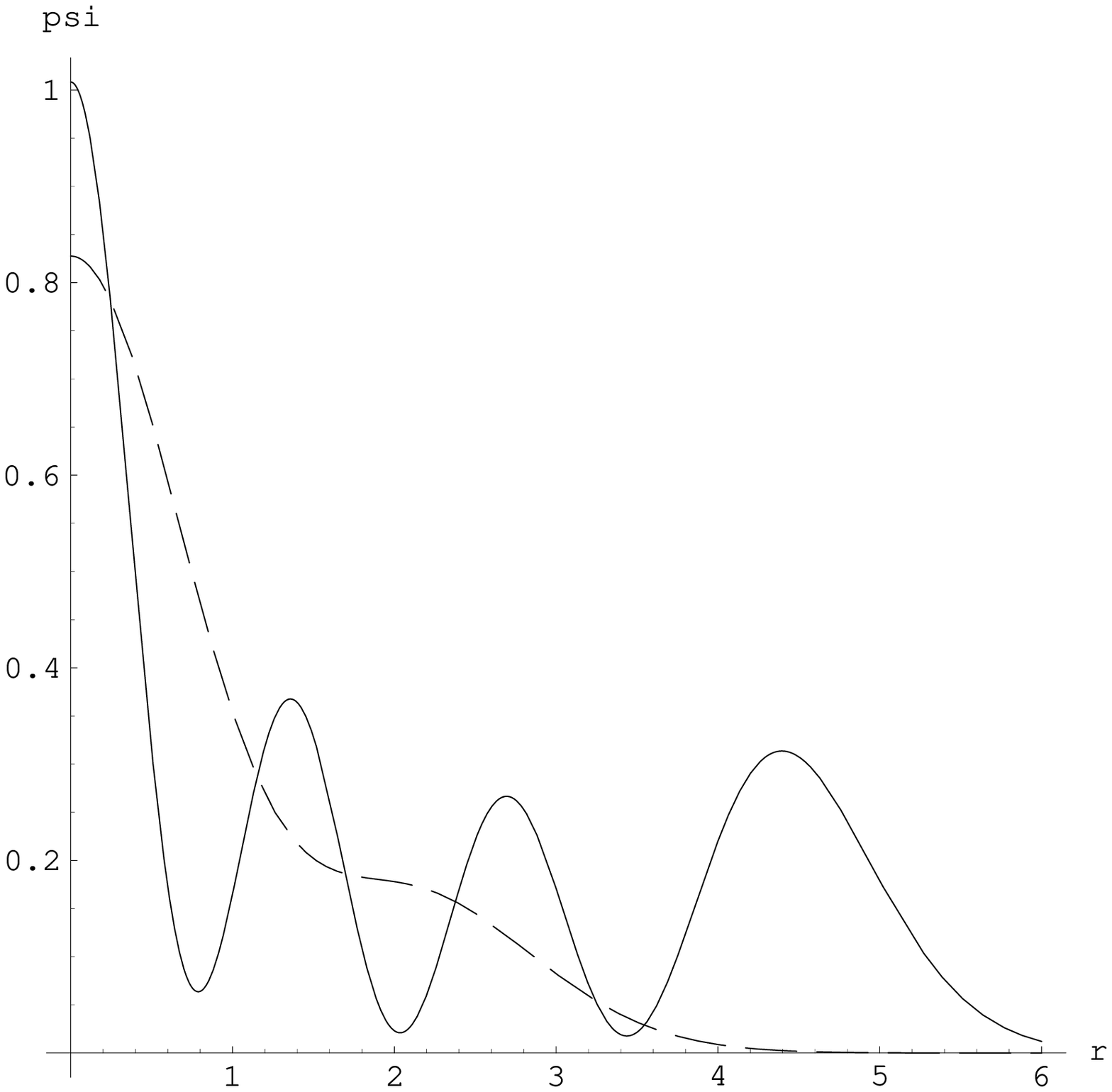}
     }
\hfill{ }
\mbox{  \epsfysize=100mm
         \epsffile{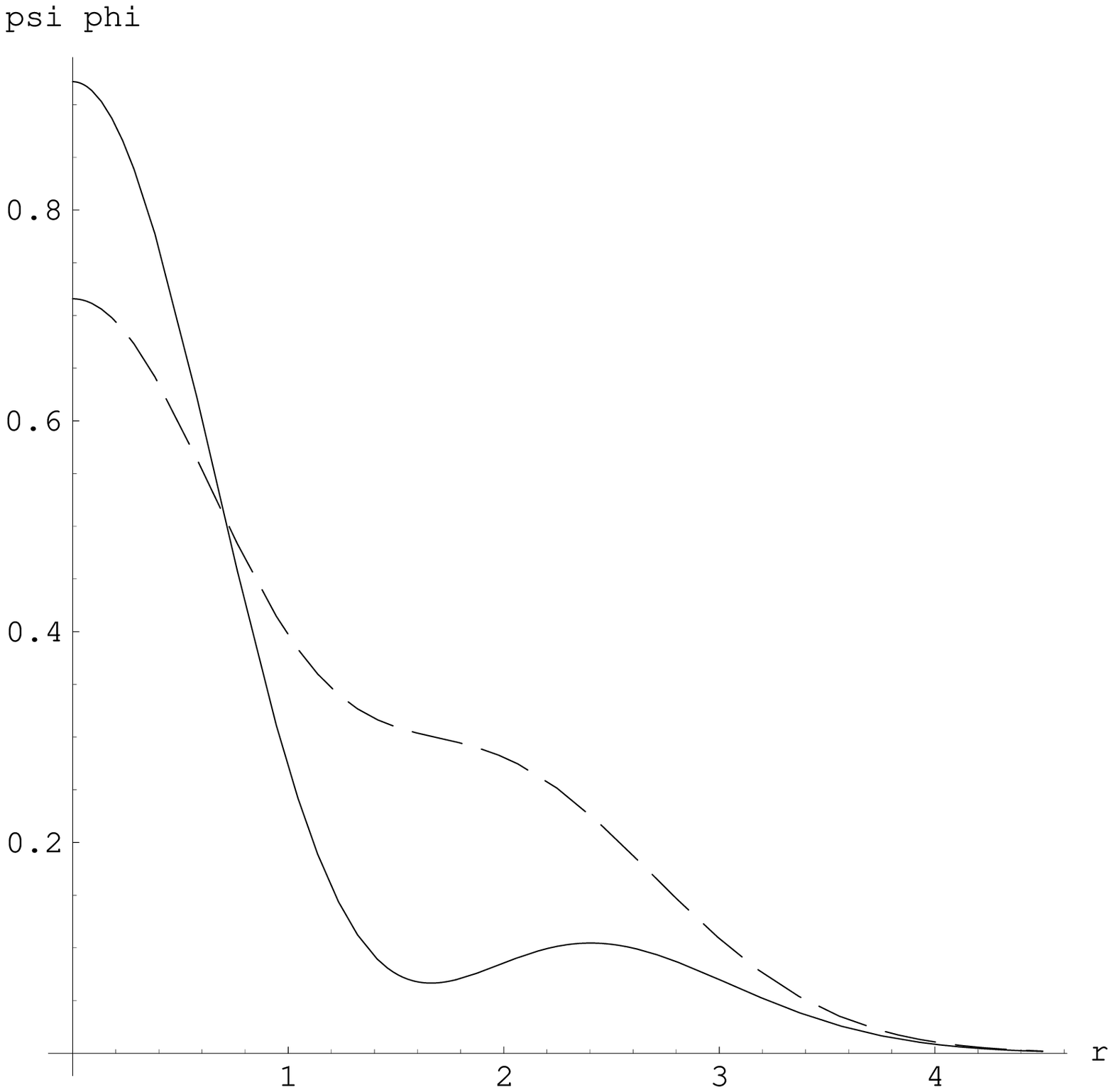}
     }
\caption{Left: two examples of self Fourier states ($\varphi=\psi$). Right:
mixed parity case ($\varphi \ne \psi$).}
\end{figure}

Some qualitative considerations may be drawn about the solutions we found in 
various examples. In the left part of Figure 12, we display two typical 
solutions with self-Fourier properties. In the right part of Fig. 12 we 
display one solution connecting two distinct partners $\psi$ and $\varphi.$ 
It is interesting to note that the shapes show distinctive features.  One 
class of shapes, which are monotonic, seem to remain closer to the bare 
Gaussian. Indeed, the bare Gaussian is the building block of our method; it 
always belongs to the subsets we found. The other class, with oscillations, 
is different.One might be concerned with the possibility that our truncations 
to a maximum order $N$ limit the flexibility of our method to the vicinity of 
the Gaussian. It seems not to be so, as shown for instance by the very 
oscillating solution\footnote{
Its equation reads \vspace{-2mm}
\begin{eqnarray*}
\psi=\exp(-r^2/2) \left[ .566053 + .0488517 (3 - 12 r^2 + 4 r^4) + 
.0011871 (105 - 840 r^2 + 840 r^4 - 224 r^6  + 16 r^8) + \right. \\ \left.
.0000164538 (10395 - 124740 r^2  + 207900 r^4  - 110880 r^6  + 23760 r^8  - 
2112 r^{10} + 64 r^{12}) \right]\, . \end{eqnarray*}
} in the left part of Fig. 12, 
which exhibits an approximate periodicity in an interval; it is reminiscent of 
the Dirac comb, which is of course outside the set of functions constructed 
with a finite number of polynomials.

Our bases are flexible enough to reconstruct any function having positivity 
properties, but in some cases convergence might be slow. It is not excluded 
that other bases exist, that might be more convenient to speed up the 
convergence and make easier the search for positivity domains. Another open
problem is that of positivity for periodic functions. Such questions are 
beyond the scope of the present paper.

\begin{acknowledgments}
It is a pleasure to thank R. Balian, R. Enberg, R. Lacaze, C. Marquet, 
P. Moussa, G. Soyez and A. Voros for stimulating discussions. R.P. thanks 
T. Lafforgue,  (Lyc\'ee Blaise Pascal, Orsay), for his fruitful contributions.
\end{acknowledgments}


\begin{thebibliography}{99}

\bibitem{Glau} R.J. Glauber, in {\em High-Enery Physics and Nuclear 
Structure}, Proc. of the 2nd International Conference, Rehovoth, 1967, 
ed. G. Alexander, North-Holland, Amsterdam, 1967, p.311, and the references 
mentioned therein.

\bibitem{Kovchegov}
Y.~V. Kovchegov,
\newblock Phys. Rev. {\bf D60}, 034008 (1999); Phys. Rev. {\bf D61}, 074018
(2000).

\bibitem{levy}
P.\ L\'evy: Fonctions caract\'eristiques positives [Positive 
characteristic functions]. {\it Comptes Rendus Hebdomadaires
des S\'eances de l'Acad\'emie des Sciences, S\'erie A,
Sciences Math\'ematiques} {\bf 265} (1967) 249-252. [in French]
Reprinted in: {\it \OE uvres de Paul L\'evy. Volume III.
El\'ements Al\'eatoires}, edited by
D.\ Dugu\'e in collaboration with  P.\ Deheuvels, M.\ Ib\'ero.
Gauthier-Villars \'Editeur, Paris, 1976, pp.\ 607-610.

\bibitem{Titch}
E.C. Titchmarsh (1959), {\it 
Introduction to the theory of Fourier integrals},
Clarendon Press, Oxford.

\bibitem{Caola:1991ad}
M.~J.~Caola,
J.\ Phys.\ A {\bf 24}, L1143 (1991). 

\bibitem{Cincotti:1991ad}
G.~Cincotti, F.~ Gori, M.~Santarsiero, 
J.\ Phys.\ A {\bf 25}, L1191 (1992). 

\bibitem{lafforgue} T. Lafforgue, private communication.

\bibitem{Scharnhorst:2002im}
For a comprehensive list of references, see  K.~Scharnhorst,
J.\ Math.\ Phys.\  {\bf 44}, 5415 (2003)
[arXiv:math-ph/0206006], in particular the references [111] to [125].

\bibitem{gelfand}
I.M. Gel'fand and N.Ya. Vilenkin (1968), {\it Generalized Functions}, Vol.IV,
Academic Press, New-York and London.

\bibitem{Teugels:1971ge}
See, for general mathematical properties, 

\noindent J.~L.~Teugels,
``Probability Density Functions Which Are Their Own Characteristic 
Functions,''
%
Bull.\ Soc.\ Math.\ Belg.\  {\bf 23}, 236 (1971);

\noindent H.-J.\ Rossberg: ``Positive definite probability densities
and probability distributions,''
{\it Journal of Mathematical Sciences (New York)} 
{\bf 76} (1995) 2181-2197. This article is part of: 
Problemy Usto\u{\i}chivosti 
Stokhasticheskikh Modele\u{\i} [Problems of the Stability
of Stochastic Models] -- Trudy Seminara (Proceedings
of the Seminar on Stochastic Problems), Moscow, 1993.
{\it Journal of Mathematical Sciences (New York)} 
{\bf 76}:1 (1995) 2093--2220;

\noindent
K.~Schladitz and H.~J.~Engelbert,
 ``On probability density functions which are their own characteristic
functions,''
Teor.\ Veroyatn.\ Primen.\  {\bf 40}, 694 (1995)
[Theor.\ Probab.\ Appl.\  {\bf 40}, 577 (1996)].

\bibitem{sturm} C. Sturm, ``M\'emoire sur la r\'esolution des \'equations 
num\'eriques'', Bull. Sci. Math. Ferussac {\bf 2} (1829), 419 [in French].

\end{thebibliography}
\end{document}